# Improved Bounds on the Parity-Check Density and Achievable Rates of Binary Linear Block Codes with Applications to LDPC Codes

Gil Wiechman    Igal Sason

Technion – Israel Institute of Technology
Haifa 32000, Israel
{igillw@tx, sason@ee}.technion.ac.il

June 27, 2018

## Abstract

We derive bounds on the asymptotic density of parity-check matrices and the achievable rates of binary linear block codes transmitted over memoryless binary-input output-symmetric (MBIOS) channels. The lower bounds on the density of arbitrary parity-check matrices are expressed in terms of the gap between the rate of these codes for which reliable communication is achievable and the channel capacity, and the bounds are valid for every sequence of binary linear block codes. These bounds address the question, previously considered by Sason and Urbanke, of how sparse can parity-check matrices of binary linear block codes be as a function of the gap to capacity. Similarly to a previously reported bound by Sason and Urbanke, the new lower bounds on the parity-check density scale like the log of the inverse of the gap to capacity, but their tightness is improved (except for a binary symmetric/erasure channel, where they coincide with the previous bound). The new upper bounds on the achievable rates of binary linear block codes tighten previously reported bounds by Burshtein *et al.*, and therefore enable to obtain tighter upper bounds on the thresholds of sequences of binary linear block codes under ML decoding. The bounds are applied to low-density parity-check (LDPC) codes, and the improvement in their tightness is exemplified numerically. The upper bounds on the achievable rates enable to assess the inherent loss in performance of various iterative decoding algorithms as compared to optimal ML decoding. The lower bounds on the asymptotic parity-check density are helpful in assessing the inherent tradeoff between the asymptotic performance of LDPC codes and their decoding complexity (per iteration) under message-passing decoding.

## 1    Introduction

Error correcting codes which employ iterative decoding algorithms are now considered state of the art in the field of low-complexity coding techniques. The graphical representation of these codes is used to describe their algebraic structure, and also enables a unified description of their iterative decoding algorithms over various channels. These codes closely approach the capacity limit of many standard communication channels under iterative decoding. By now, there is a large collection of families of iteratively decoded codes including low-density parity-check (LDPC), turbo, repeat-accumulate and product codes; all of them, demonstrate a rather small gap (in rate) to capacity with feasible complexity. In [6], Khandekar and McEliece have suggested to study the encoding and decoding complexities of ensembles of iteratively decoded codes on graphs as a function of their gap to capacity. They conjectured that if the achievable rate under iterative message-passing decoding is a fraction $1 - \varepsilon$ of the channel capacity, then for a wide class of channels, the encoding complexity scales like $\ln \frac{1}{\varepsilon}$ and the decoding complexity scales like $\frac{1}{\varepsilon} \ln \frac{1}{\varepsilon}$. The only exception is the binary erasure channel (BEC) where the decoding complexity behaves like $\ln \frac{1}{\varepsilon}$ (same as encoding complexity) because of the absolute reliability of the messages passed through the edges of the graph (hence, every edge can be used only once during the iterative decoding process).

LDPC codes are efficiently encoded and decoded due to the sparseness of their parity-check matrices. In his thesis [4], Gallager proved that right-regular LDPC codes (i.e., LDPC codes with a constant degree ($a_R$) of the parity-check nodes) cannot achieve the channel capacity on a BSC, even under optimal ML decoding. This inherent gap to capacity is well approximated by an expression which decreases to zero exponentially fast in $a_R$. Richardson *et al.* [11] have extended this result, and proved that the same conclusion holds if $a_R$ designates the *maximal right degree*. Sason and Urbanke later observed in [13] that the result still applies when considering the *average right degree*. Gallager's bound [4, Theorem 3.3] provides an upper bound on the rate of right-regular LDPC codes which achieve reliable communications over the BSC. Burshtein *et al.* have generalized Gallager's bound for a general MBIOS channel [1], and the work in [13] relies on their generalization.

Consider the number of ones in a parity-check matrix which represents a binary linear code, and normalize it per information bit (i.e., with respect to the dimension of the code). This quantity (which will be later defined as the *density* of the parity-check matrix) is equal to $\frac{1-R}{R}$ times the average right degree of the bipartite graph that represents the code, where $R$ is the rate of the code in bits per channel use. In [13], Sason and Urbanke considered how sparse can parity-check matrices of binary linear block codes be, as a function of their gap to capacity (where this gap depends in general on the channel and on the decoding algorithm). An information-theoretic lower bound on the asymptotic density of parity-check matrices was derived in [13, Theorem 2.1] where this bound applies to every MBIOS channel and *every* sequence of binary linear block codes achieving a fraction $1 - \varepsilon$ of the channel capacity with vanishing bit error probability. It holds for an arbitrary representation of parity-check matrices for these codes, and is of the form $\frac{K_1 + K_2 \ln \frac{1}{\varepsilon}}{1-\varepsilon}$ where $K_1$ and $K_2$ are constants which only depend on the channel. Though the logarithmic behavior of this lower bound is in essence correct (due to a logarithmic behavior of the upper bound on the asymptotic parity-check density in [13, Theorem 2.2]), the lower bound in [13, Theorem 2.1] is *not* tight (with the exception of the BEC, as demonstrated in [13, Theorem 2.3], and possibly also the binary symmetric channel (BSC)). The derivation of the bounds in this paper was motivated by the desire to improve the results in [1, Theorems 1 and 2] and [13, Theorem 2.1] which are based on a two-level quantization of the log-likelihood ratio (LLR).

In [7], Measson and Urbanke derived an upper bound on the maximum-likelihood (ML) thresholds of LDPC ensembles when the codes are transmitted over the BEC. Their general approach relies on EXIT-like functions and the area theorem. This bound coincides with the ML threshold determined by Montanari *et al.* using the replica method, showing that the bound is in fact tight. In [8], Montanari presented a new approach for the analysis of codes on graphs under maximum a posteriori probability (MAP) decoding. His approach is based on statistical mechanics, and the resulting expressions are related to the density evolution analysis of belief propagation decoding.

Motivated by the heuristic statistical mechanics results, it was conjectured in [8] that the bounds on the asymptotic achievable rates of LDPC codes are tight.

We derive in this paper improved bounds on the achievable rates and the asymptotic parity-check density of sequences of binary linear block codes. The bounds in [1, 13] and this paper are valid for *every* sequence of binary linear block codes, in contrast to a high probability result which was previously derived for the binary erasure channel (BEC) from density evolution analysis [14]. Shokrollahi proved in [14] that when the codes are communicated over a BEC, the growth rate of the average right degree (i.e., the average degree of the parity-check nodes in a bipartite Tanner graph) is at least logarithmic in terms of the gap to capacity. The statement in [14] is a high probability result, and hence it is not necessarily satisfied for every particular code from this ensemble. Further, it assumes a sub-optimal (iterative) decoding algorithm, where the statements in [1, 13] and this paper are valid even under optimal ML decoding.

The significance of the bounds in this paper is demonstrated in two respects. The new upper bounds on the achievable rates of binary linear block codes tighten previously reported bounds by Burshtein *et al.* [1], and therefore enable to obtain tighter upper bounds on the thresholds of sequences of binary linear block codes under ML decoding. They are applied to LDPC codes, and the improvement in their tightness is exemplified numerically. Comparing the new upper bounds on the achievable rates with thresholds provided by a density-evolution analysis gives rigorous bounds on the inherent loss in performance due to the sub-optimality of iterative message-passing decoding (as compared to soft-decision ML decoding). The new lower bounds on the asymptotic parity-check density tighten the lower bound in [13, Theorem 2.1]. Since the parity-check density can be interpreted as the complexity per iteration under iterative message-passing decoding, then tightening the reported lower bound on the parity-check density [13] gives insight on the tradeoff between the asymptotic performance and decoding complexity of LDPC codes.

In this paper, preliminary material is presented in Section 2, and the theorems are introduced and proved in Sections 3 and 4. The derivation of the bounds in Section 3 was motivated by the desire to generalize the results in [1, Theorems 1 and 2] and [13, Theorem 2.1]. A two-level quantization of the log-likelihood ratio (LLR), in essence replacing the arbitrary MBIOS channel by a physically degraded binary symmetric channel (BSC), is modified in Section 3 to a quantized channel which better reflects the statistics of the original channel (though the quantized channel is still physically degraded w.r.t. the original channel). The number of quantization levels of the LLR for the new channel is an arbitrary integer power of 2, and the calculation of these bounds is subject to an optimization of the quantization levels, as to obtain the tightest bounds within their form. In Section 4, we rely on the conditional *pdf* of the LLR of the MBIOS channel, and operate on an equivalent channel without quantizing the LLR. This second approach finally leads to bounds which are uniformly tighter than the bounds we derive in Section 3. It appears to be even simpler to calculate the un-quantized bounds in Section 4, as their calculation do not involve the solution of any optimization equation (in contrast to the quantized bounds, whose calculation involves a numerical solution of optimization equations w.r.t. the quantization levels of the LLR). Fortunately, the multi-dimensional integral obtained in the derivation of the bounds in Section 4 is transformed to a rapidly convergent infinite series of one-dimensional integrals; this issue is crucial in facilitating the calculation of the bounds in Section 4. We note that the significance of both the quantized and un-quantized bounds in Sections 3 and 4, respectively, stems from a comparison between these bounds which gives insight on the effect of the number of quantization levels of the LLR (even if they are optimally determined) on the achievable rates, as compared to the ideal case where no quantization is done. Numerical results are exemplified and explained in Section 5. Finally, we summarize our discussion in Section 6 and present interesting issues which deserve further research. Four appendices provide further technical details referring to the proofs in Sections 3 and 4.

## 2 Preliminaries

We introduce here some definitions and theorems from [1, 13] which serve as a preliminary material for the rest of the paper. Definitions 2.1 and 2.2 are taken from [13, Section 2].

**Definition 2.1 (Capacity-Approaching Codes).** Let $\{\mathcal{C}_m\}$ be a sequence of codes of rate $R_m$, and assume that for every $m$, the codewords of the code $\mathcal{C}_m$ are transmitted with equal probability over a channel whose capacity is $C$. This sequence is said to *achieve a fraction $1-\varepsilon$ of the channel capacity with vanishing bit (block) error probability* if $\lim_{m\to\infty} R_m = (1-\varepsilon)C$, and if there exists a decoding algorithm under which the average bit (block) error probability of the code $\mathcal{C}_m$ tends to zero in the limit where $m \to \infty$.

**Definition 2.2 (Parity-Check Density).** Let $\mathcal{C}$ be a binary linear code of rate $R$ and block length $n$, which is represented by a parity-check matrix $H$. We define the *density* of $H$, call it $\Delta = \Delta(H)$, as the normalized number of ones in $H$ *per information bit*. The total number of ones in $H$ is therefore equal to $nR\Delta$.

**Definition 2.3 (Log-Likelihood Ratio (LLR)).** Let $p_{Y|X}(\cdot|\cdot)$ be the conditional *pdf* of an arbitrary MBIOS channel. The log-likelihood ratio (LLR) at the output of the channel is

$$\mathrm{LLR}(y) \triangleq \ln\left(\frac{p_{Y|X}(y|X=0)}{p_{Y|X}(y|X=1)}\right).$$

Throughout the paper, we assume that all the codewords of a binary linear block code are equally likely to be transmitted. Also, we use the notation $h_2(\cdot)$ for the binary entropy function to base 2, i.e., $h_2(\cdot) = -x\log_2(x) - (1-x)\log_2(1-x)$.

**Theorem 2.1 (An Upper Bound on the Achievable Rates for Reliable Communication over MBIOS Channels).** [1, Theorem 2]: Consider a sequence $\{\mathcal{C}_m\}$ of binary linear block codes of rate $R_m$, and assume that their block length tends to infinity as $m \to \infty$. Let $H_m$ be a parity-check matrix of the code $\mathcal{C}_m$, and assume that $d_{k,m}$ designates the fraction of the parity-check equations involving $k$ variables. Let

$$d_k \triangleq \liminf_{m\to\infty} d_{k,m}, \quad R \triangleq \liminf_{m\to\infty} R_m. \tag{1}$$

Suppose that the transmission of these codes takes place over an MBIOS channel with capacity $C$ bits per channel use, and let

$$w \triangleq \frac{1}{2}\int_{-\infty}^{\infty} \min\bigl(f(y), f(-y)\bigr)\,dy \tag{2}$$

where $f(y) \triangleq p_{Y|X}(y|X=1)$ designates the conditional *pdf* of the output of the MBIOS channel. Then, a necessary condition for vanishing block error probability as $m \to \infty$ is

$$R \leq 1 - \frac{1-C}{\sum_k \left\{ d_k\, h_2\left(\frac{1-(1-2w)^k}{2}\right)\right\}}.$$

**Theorem 2.2 (Lower Bounds on the Asymptotic Parity-Check Density with 2-Levels Quantization).** [13, Theorem 2.1]: Let $\{\mathcal{C}_m\}$ be a sequence of binary linear codes achieving a fraction $1-\varepsilon$ of the capacity of an MBIOS channel with vanishing *bit error probability*. Then, the asymptotic density $(\Delta_m)$ of their parity-check matrices satisfies

$$\liminf_{m\to\infty} \Delta_m > \frac{K_1 + K_2 \ln\frac{1}{\varepsilon}}{1-\varepsilon} \tag{3}$$

where

$$K_1 = \frac{(1-C) \ln\left(\frac{1}{2\ln 2} \frac{1-C}{C}\right)}{2C \ln\left(\frac{1}{1-2w}\right)}, \quad K_2 = \frac{1-C}{2C \ln\left(\frac{1}{1-2w}\right)} \tag{4}$$

and $w$ is defined in (2). For a BEC with erasure probability $p$, the coefficients $K_1$ and $K_2$ in (4) are improved to

$$K_1 = \frac{p \ln\left(\frac{p}{1-p}\right)}{(1-p) \ln\left(\frac{1}{1-p}\right)}, \quad K_2 = \frac{p}{(1-p) \ln\left(\frac{1}{1-p}\right)}. \tag{5}$$

Using standard notation, an ensemble of $(n, \lambda, \rho)$ LDPC codes is characterized by its length $n$, and the polynomials $\lambda(x) = \sum_{i=2}^{\infty} \lambda_i x^{i-1}$ and $\rho(x) = \sum_{i=2}^{\infty} \rho_i x^{i-1}$, where $\lambda_i$ ($\rho_i$) is equal to the probability that a randomly chosen edge is connected to a variable (parity-check) node of degree $i$. The variables (parity-check sets) are represented by the left (right) nodes of a bipartite graph which represents an LDPC code.

## 3 Approach I: Bounds Based on Quantization of the LLR

In this section, we introduce bounds on the achievable rates and the asymptotic parity-check density of sequences of binary linear block codes. The bounds generalize previously reported results in [1] and [13] which were based on a symmetric two-level quantization of the LLR. This is achieved by extending the concept of quantization to an arbitrary integer power of 2; to this end, our analysis relies on the Galois field $\text{GF}(2^m)$. In Section 3.1, we demonstrate the results and their proofs for four-level quantization. In Section 3.2, we extend the results to a symmetric quantization with a number of levels which is an arbitrary integer power of 2. This order of presentation was chosen since many concepts which are helpful for the generalization in Section 3.2 are written in a simplified notation for the four-level quantization, along with all the relevant lemmas for the general case which are already introduced in the derivation of the bound with four-level quantization. This also shortens considerably the proof for the general quantization in Section 3.2.

### 3.1 Bounds for Four-Levels of Quantization

As a preparatory step towards developing bounds on the parity-check density and the rate of binary linear block codes, we present a lower bound on the conditional entropy of a transmitted codeword given the received sequence at the output of an arbitrary MBIOS channel.

**Proposition 3.1.** Let $\mathcal{C}$ be a binary linear block code of length $n$ and rate $R$. Let $\mathbf{x} = (x_1, \ldots x_n)$ and $\mathbf{y} = (y_1, \ldots, y_n)$ designate the transmitted codeword and received sequence, respectively, when the communication takes place over an MBIOS channel with conditional pdf $p_{Y|X}(\cdot|\cdot)$. For an arbitrary positive $l \in \mathbb{R}^+$, let us define the probabilities $p_0, p_1, p_2, p_3$ as follows:

$$p_0 \triangleq \Pr\{\text{LLR}(Y) > l \mid X = 0\}$$
$$p_1 \triangleq \Pr\{\text{LLR}(Y) \in (0, l] \mid X = 0\} + \frac{1}{2} \Pr\{\text{LLR}(Y) = 0 \mid X = 0\}$$
$$p_2 \triangleq \Pr\{\text{LLR}(Y) \in [-l, 0) \mid X = 0\} + \frac{1}{2} \Pr\{\text{LLR}(Y) = 0 \mid X = 0\}$$
$$p_3 \triangleq \Pr\{\text{LLR}(Y) < -l \mid X = 0\}. \tag{6}$$

For an arbitrary parity-check matrix of the code $\mathcal{C}$, let $d_k$ designate the fraction of the parity-checks involving $k$ variables. Then, the conditional entropy of the transmitted codeword given the received sequence satisfies

$$\frac{H(\mathbf{X}|\mathbf{Y})}{n} \geq 1 - C - (1-R) \cdot$$

$$\cdot \sum_k \left\{ d_k \sum_{t=0}^{k} \binom{k}{t} (p_1+p_2)^t (p_0+p_3)^{k-t} h_2 \left( \frac{1 - \left(1 - \frac{2p_2}{p_1+p_2}\right)^t \left(1 - \frac{2p_3}{p_0+p_3}\right)^{k-t}}{2} \right) \right\} . (7)$$

*Proof.* Considering an MBIOS channel whose conditional *pdf* is given by $p_{Y|X}(\cdot|\cdot)$, we introduce a new physically degraded channel. It is a binary-input, quaternary-output symmetric channel (see Fig. 1). To this end, let $l \in \mathbb{R}^+$ be an arbitrary positive number, and let $\alpha$ be a primitive element of the Galois field $\mathrm{GF}(2^2)$ (so $\alpha^2 = 1 + \alpha$). The set of the elements of this field is $\{0, 1, \alpha, 1+\alpha\}$. Let $X_i$ and $Y_i$ designate the random variables referring to the input and output of the original channel $p_{Y|X}(\cdot|\cdot)$ at time $i$ (where $i = 1, 2, \ldots, n$). We define the degraded channel as a channel with four quantization levels of the LLR. The output of the degraded channel at time $i$, $z_i$, is calculated from the output $y_i$ of the original channel as follows:

- If $\mathrm{LLR}(y_i) > l$, then $z_i = 0$.
- If $0 < \mathrm{LLR}(y_i) \leq l$, then $z_i = \alpha$.
- If $-l \leq \mathrm{LLR}(y_i) < 0$, then $z_i = 1 + \alpha$.
- If $\mathrm{LLR}(y_i) < -l$, then $z_i = 1$.
- If $\mathrm{LLR}(y_i) = 0$, then $z_i$ is chosen as $\alpha$ or $1+\alpha$ with equal probability ($\frac{1}{2}$).

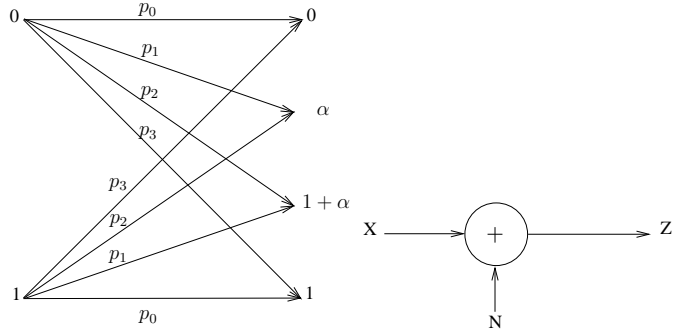

Figure 1: The channel model in the left plot is a physically degraded channel used for the derivation of the bound with four levels of quantization. The element $\alpha$ denotes a primitive element in $\mathrm{GF}(2^2)$. This channel model is equivalent to a channel with an additive noise in $\mathrm{GF}(2^2)$ (see right plot).

From the definition of the degraded channel in Fig. 1, this channel has an additive noise and is also binary-input output-symmetric. It follows that the transition probabilities given in (6) can be expressed in an equivalent way by

$$\begin{aligned} p_0 &= \Pr(Z=0 \mid X=0) = \Pr(Z=1 \mid X=1) \\ p_1 &= \Pr(Z=\alpha \mid X=0) = \Pr(Z=1+\alpha \mid X=1) \\ p_2 &= \Pr(Z=1+\alpha \mid X=0) = \Pr(Z=\alpha \mid X=1) \\ p_3 &= \Pr(Z=1 \mid X=0) = \Pr(Z=0 \mid X=1) \end{aligned}$$

where the symmetry in these transition probabilities holds since the original channel is MBIOS.

Since $\mathcal{C}$ is a binary linear block code of length $n$ and rate $R$, and the codewords are transmitted with equal probability then
$$H(\mathbf{X}) = nR. \tag{8}$$
Also, since the channel $P_{Y|X}(\cdot|\cdot)$ is memoryless, then
$$H(\mathbf{Y}|\mathbf{X}) = nH(Y|X). \tag{9}$$

We designate the output sequences of the original channel and its degraded version by $\mathbf{Y}$ and $\mathbf{Z}$, respectively. Since the mapping from $Y_i$ to the degraded output $Z_i$ ($i = 1, 2, \cdots, n$) is memoryless, then $H(\mathbf{Z}|\mathbf{Y}) = nH(Z|Y)$, and
$$\begin{aligned} H(\mathbf{Y}) &= H(\mathbf{Z}) - H(\mathbf{Z}|\mathbf{Y}) + H(\mathbf{Y}|\mathbf{Z}) \\ &= H(\mathbf{Z}) - nH(Z|Y) + H(\mathbf{Y}|\mathbf{Z}) \end{aligned} \tag{10}$$

$$\begin{aligned} H(\mathbf{Y}|\mathbf{Z}) &\leq \sum_{i=1}^n H(Y_i|Z_i) \\ &= nH(Y|Z) \\ &= n\left[H(Y) - H(Z) + H(Z|Y)\right]. \end{aligned} \tag{11}$$

Applying the above towards a lower bound on the conditional entropy $H(\mathbf{X}|\mathbf{Y})$, we get
$$\begin{aligned} H(\mathbf{X}|\mathbf{Y}) &= H(\mathbf{X}) + H(\mathbf{Y}|\mathbf{X}) - H(\mathbf{Y}) \\ &= nR + nH(Y|X) - H(\mathbf{Y}) \\ &= nR + nH(Y|X) - H(\mathbf{Z}) - H(\mathbf{Y}|\mathbf{Z}) + nH(Z|Y) \\ &\geq nR + nH(Y|X) - H(\mathbf{Z}) - n\left[H(Y) - H(Z) + H(Z|Y)\right] + nH(Z|Y) \\ &= nR - H(\mathbf{Z}) + nH(Z) - n\left[H(Y) - H(Y|X)\right] \\ &= nR - H(\mathbf{Z}) + nH(Z) - nI(X;Y) \\ &\geq nR - H(\mathbf{Z}) + nH(Z) - nC \end{aligned} \tag{12}$$
where the second equality relies on (8) and (9), the third equality relies on (10), the first inequality relies on (11), and $I(X;Y) \leq C$ is used for the last transition (where $C$ designates the capacity of the non-degraded channel). In order to obtain a lower bound on $H(\mathbf{X}|\mathbf{Y})$ from (12), we will calculate the exact entropy of the random variable $Z$, and find an upper bound on the entropy of the random vector $\mathbf{Z}$. This will finally provide the lower bound in (7).

Since $\mathcal{C}$ is a binary linear block code, then the input $X$ is equally likely to be zero or one. The output $Z$ of the degraded channel in Fig. 1 has the following probability law:
$$\begin{aligned} \Pr(Z = 0) &= \Pr(Z = 0, X = 0) + \Pr(Z = 0, X = 1) \\ &= \Pr(Z = 0 \mid X = 0)\Pr(X = 0) + \Pr(Z = 0 \mid X = 1)\Pr(X = 1) \\ &= \frac{p_0 + p_3}{2} \end{aligned}$$
and in a similar manner,
$$\Pr(Z = 1) = \frac{p_0 + p_3}{2}, \quad \Pr(Z = \alpha) = \Pr(Z = 1 + \alpha) = \frac{p_1 + p_2}{2}.$$

The entropy of the random variable $Z$ is therefore equal to

$$\begin{aligned} H(Z) &= 2\left(\frac{p_0+p_3}{2}\right)\log_2\left(\frac{2}{p_0+p_3}\right) + 2\left(\frac{p_1+p_2}{2}\right)\log_2\left(\frac{2}{p_1+p_2}\right) \\ &= 1 + (p_0+p_3)\log_2\left(\frac{1}{p_0+p_3}\right) + (p_1+p_2)\log_2\left(\frac{1}{p_1+p_2}\right) \\ &= 1 + h_2(p_1+p_2) \end{aligned} \quad (13)$$

where the last transition follows from the equality $p_0 + p_1 + p_2 + p_3 = 1$. We now derive an upper bound on the entropy $H(\mathbf{Z})$. To this end, let

$$Z_i = \Theta_i + \Phi_i\,\alpha, \quad i = 1, 2, \ldots, n \quad (14)$$

where $\boldsymbol{\Theta} = (\Theta_1, \Theta_2, \ldots, \Theta_n)$ and $\boldsymbol{\Phi} = (\Phi_1, \Phi_2, \ldots, \Phi_n)$ are random vectors over $\{0,1\}^n$. From the composition of $Z_i$ into a pair of two binary components $(\Theta_i, \Phi_i)$, it follows from Fig. 1 that

$$\Pr(\Phi_i = 0) = p_1 + p_2, \quad \Pr(\Phi_i = 1) = p_0 + p_3 = 1 - (p_1 + p_2), \quad i = 1, 2, \ldots, n. \quad (15)$$

Based on (14) and (15), it is easy to verify the following chain of equalities:

$$\begin{aligned} H(\mathbf{Z}) &= H(Z_1, Z_2, \ldots, Z_n) \\ &= H(\Theta_1, \Theta_2, \ldots, \Theta_n, \Phi_1, \Phi_2, \ldots, \Phi_n) \\ &= H(\Phi_1, \Phi_2, \ldots, \Phi_n) + H(\Theta_1, \Theta_2, \ldots, \Theta_n \mid \Phi_1, \Phi_2, \ldots, \Phi_n) \\ &= n\,h_2(p_1+p_2) + H(\Theta_1, \Theta_2, \ldots, \Theta_n \mid \Phi_1, \Phi_2, \ldots, \Phi_n) \end{aligned} \quad (16)$$

where the last equality follows from (15) and since the degraded channel in Fig. 1 is memoryless. Let us define the syndrome at the output of the degraded channel as

$$\mathbf{S} \triangleq (\Theta_1, \Theta_2, \ldots, \Theta_n)\,H^T$$

where $H$ is a parity-check matrix of the binary linear block code $\mathcal{C}$. We note that the calculation of the syndrome only takes into account the first components in the composition of the vector $Z$ in (14). Note that the transmitted codeword $\mathbf{x} \in \mathcal{C}$ only affects the $\Theta$-components of the vector $\mathbf{Z}$ in (14), and also $\mathbf{x}H^T = 0$ for any such a codeword. Let us define $L$ as the index of the vector $(\Theta_1, \Theta_2, \ldots, \Theta_n)$ in the coset referring to the syndrome $\mathbf{S}$. Since each coset has exactly $2^{nR}$ elements which are equally likely, then $H(L) = nR$, and

$$\begin{aligned} H(\Theta_1, \Theta_2, \ldots, \Theta_n \mid \Phi_1, \Phi_2, \ldots, \Phi_n) &= H(\mathbf{S}, L \mid \Phi_1, \Phi_2, \ldots, \Phi_n) \\ &\leq H(L) + H(\mathbf{S} \mid \Phi_1, \Phi_2, \ldots, \Phi_n) \\ &= nR + H(\mathbf{S} \mid \Phi_1, \Phi_2, \ldots, \Phi_n). \end{aligned} \quad (17)$$

Considering a parity-check equation involving $k$ variables, let $\{i_1, i_2, \ldots, i_k\}$ be the set of indices of the variables involved in this parity-check equation. The relevant component of the syndrome $\mathbf{S}$ which refers to this parity-check equation is equal to zero or one if and only if the components of the sub-vector $(\Theta_{i_1}, \Theta_{i_2}, \ldots, \Theta_{i_k})$ differ from the components of the sub-vector $(X_{i_1}, X_{i_2}, \ldots, X_{i_k})$ in an even or odd number of indices, respectively. It is clear from Fig. 1 that for an index $i$ for which $\Phi_i = 1$, the random variables $X_i$ and $\Theta_i$ are different in probability $\frac{p_2}{p_1+p_2}$; as a result of the symmetry of the channel, this probability is independent of the value of $X_i$. Similarly, for an index $i$ for which $\Phi_i = 0$, the random variables $X_i$ and $\Theta_i$ are different in probability $\frac{p_3}{p_0+p_3}$, which again is independent of the value of $X_i$.

Given that the Hamming weight of the vector $(\Phi_{i_1}, \Phi_{i_2}, \ldots, \Phi_{i_k})$ is $t$, then the probability that the components of the two random vectors $(\Theta_{i_1}, \Theta_{i_2}, \ldots, \Theta_{i_k})$ and $(X_{i_1}, X_{i_2}, \ldots, X_{i_k})$ differ an even number of times is equal to

$$q_1(t,k)\, q_2(t,k) + \big(1 - q_1(t,k)\big)\big(1 - q_2(t,k)\big)$$

where $q_1(t,k)$ designates the probability that among the $t$ indices $i$ for which $\Phi_i = 1$, the random variables $X_i$ and $\Theta_i$ differ an even number of times, and $q_2(t,k)$ designates the probability that among the $k-t$ indices $i$ for which $\Phi_i = 0$, the random variables $X_i$ and $\Theta_i$ differ an even number of times. Based on the discussion above, it follows that

$$q_1(t,k) = \sum_{i \text{ even}} \binom{t}{i}\left(\frac{p_2}{p_1+p_2}\right)^i = \frac{1 + \left(1 - \frac{2p_2}{p_1+p_2}\right)^t}{2}$$

$$q_2(t,k) = \sum_{i \text{ even}} \binom{k-t}{i}\left(\frac{p_3}{p_0+p_3}\right)^i = \frac{1 + \left(1 - \frac{2p_3}{p_0+p_3}\right)^{k-t}}{2}$$

so the probability that the two vectors $(\Theta_{i_1}, \Theta_{i_2}, \ldots, \Theta_{i_k})$ and $(X_{i_1}, X_{i_2}, \ldots, X_{i_k})$ differ in an even number of indices is

$$q_1(t,k)\,q_2(t,k) + \big(1 - q_1(t,k)\big)\big(1 - q_2(t,k)\big) = \frac{1 + \left(1 - \frac{2p_2}{p_1+p_2}\right)^t \left(1 - \frac{2p_3}{p_0+p_3}\right)^{k-t}}{2}.$$

We conclude that given a vector $\underline{\Phi} \in \{0,1\}^k$ of Hamming weight $t$

$$H\big(S_i \mid (\Phi_{i_1}, \Phi_{i_2}, \ldots, \Phi_{i_k}) = \underline{\Phi}\big) = h_2\left(\frac{1 + \left(1 - \frac{2p_2}{p_1+p_2}\right)^t \left(1 - \frac{2p_3}{p_0+p_3}\right)^{k-t}}{2}\right).$$

This yields that if the calculation of a component $S_i$ $(i = 1, \ldots, n(1-R))$ in the syndrome $\mathbf{S}$ relies on a parity-check equation involving $k$ variables, then

$$
\begin{aligned}
H(S_i \mid \Phi_1, \Phi_2, \ldots, \Phi_n) &= H(S_i \mid \Phi_{i_1}, \Phi_{i_2}, \ldots, \Phi_{i_k}) \\
&= \sum_{\underline{\Phi} \in \{0,1\}^k} \Pr\big((\Phi_{i_1}, \Phi_{i_2}, \ldots, \Phi_{i_k}) = \underline{\Phi}\big) \cdot H\big(S_i \mid (\Phi_{i_1}, \Phi_{i_2}, \ldots, \Phi_{i_k}) = \underline{\Phi}\big) \\
&= \sum_{t=0}^{k} \binom{k}{t}(p_1+p_2)^t(p_0+p_3)^{k-t} h_2\left(\frac{1 + \left(1 - \frac{2p_2}{p_1+p_2}\right)^t \left(1 - \frac{2p_3}{p_0+p_3}\right)^{k-t}}{2}\right) \\
&= \sum_{t=0}^{k} \binom{k}{t}(p_1+p_2)^t(p_0+p_3)^{k-t} h_2\left(\frac{1 - \left(1 - \frac{2p_2}{p_1+p_2}\right)^t \left(1 - \frac{2p_3}{p_0+p_3}\right)^{k-t}}{2}\right)
\end{aligned}
$$

where the third equality turns to averaging over the Hamming weight of $\underline{\Phi} = (\Phi_{i_1}, \Phi_{i_2}, \ldots, \Phi_{i_k})$, and the last equality follows from the symmetry of the binary entropy function (where $h_2(x) = h_2(1-x)$ for $x \in [0,1]$). Let $d_k$ designate the fraction of parity-check equations in the arbitrary parity-check matrix which involve $k$ variables, so their total number is $n(1-R)d_k$ and

$$
\begin{aligned}
&H(\mathbf{S} \mid \Phi_1, \Phi_2, \ldots, \Phi_n) \\
&\leq \sum_{i=1}^{n(1-R)} H(S_i \mid \Phi_1, \Phi_2, \ldots, \Phi_n) \\
&= n(1-R) \sum_{k} \left\{ d_k \sum_{t=0}^{k} \binom{k}{t}(p_1+p_2)^t(p_0+p_3)^{k-t} h_2\left(\frac{1 - \left(1 - \frac{2p_2}{p_1+p_2}\right)^t \left(1 - \frac{2p_3}{p_0+p_3}\right)^{k-t}}{2}\right) \right\}.
\end{aligned}
\tag{18}
$$

By combining (16)–(18), an upper bound on the entropy of the random vector **Z** follows:

$$H(\mathbf{Z}) \leq nR + nh_2(p_1 + p_2)$$

$$+ n(1-R) \sum_k \left\{ d_k \sum_{t=0}^{k} \binom{k}{t} (p_1+p_2)^t (p_0+p_3)^{k-t} h_2 \left( \frac{1 - \left(1 - \frac{2p_2}{p_1+p_2}\right)^t \left(1 - \frac{2p_3}{p_0+p_3}\right)^{k-t}}{2} \right) \right\}. \quad (19)$$

The substitution of (13) and (19) in (12) finally provides the lower bound on the conditional entropy $H(\mathbf{X} \mid \mathbf{Y})$ in (7). □

The following theorem tightens the lower bound on the parity-check density of an arbitrary sequence of binary linear block codes given in [13, Theorem 2.1]:

**Theorem 3.1 ("Four-Level Quantization" Lower Bound on the Asymptotic Parity-Check Density of Binary Linear Block Codes).** Let $\{C_m\}$ be a sequence of binary linear block codes achieving a fraction $1 - \varepsilon$ of the capacity of an MBIOS channel with vanishing *bit error probability*. Then, the asymptotic density $(\Delta_m)$ of their parity-check matrices satisfies

$$\liminf_{m \to \infty} \Delta_m > \frac{K_1 + K_2 \ln \frac{1}{\varepsilon}}{1 - \varepsilon} \quad (20)$$

where

$$K_1 = K_2 \ln \left( \frac{1}{2\ln(2)} \frac{1-C}{C} \right), \quad K_2 = -\frac{1-C}{C \ln \left( \frac{(p_1-p_2)^2}{(p_1+p_2)} + \frac{(p_0-p_3)^2}{p_0+p_3} \right)} \quad (21)$$

and $p_0, p_1, p_2, p_3$ are defined in (6) in terms of $l \in \mathbb{R}^+$. The optimal value of $l$ is given implicitly by the equation

$$\frac{p_2^2 + e^{-l} p_1^2}{(p_1+p_2)^2} = \frac{p_3^2 + e^{-l} p_0^2}{(p_0+p_3)^2} \quad (22)$$

where such a solution always exists.[1]

*Proof.* Derivation of the lower bound in (20) and (21):

**Lemma 3.1.** Let $\mathcal{C}$ be a binary linear block code of length $n$ and rate $R$. Let $P_b$ designate the average bit error probability of the code $\mathcal{C}$ which is associated with an arbitrary decoding algorithm and channel, and let $\mathbf{X}$ and $\mathbf{Y}$ designate the transmitted codeword and received sequence, respectively. Then

$$\frac{H(\mathbf{X} \mid \mathbf{Y})}{n} \leq R \, h_2(P_b). \quad (23)$$

*Proof.* The lemma is proved in Appendix A.1. □

**Lemma 3.2.** $h_2(x) \leq 1 - \frac{2}{\ln 2} \left(\frac{1}{2} - x\right)^2$ for $0 \leq x \leq 1$.

*Proof.* The lemma is proved in [13, Lemma 3.1]. □

---
[1] It was observed numerically that the solution $l$ of the optimization equation (22) is unique when considering the binary-input AWGN channel. We conjecture that the uniqueness of such a solution is a property which holds for MBIOS channels under some mild conditions.

Referring to an arbitrary sequence of binary linear block codes $\{\mathcal{C}_m\}$ which achieves a fraction $1-\varepsilon$ to capacity with vanishing bit error probability, then according to Definition 2.1, there exists a decoding algorithm (e.g., ML decoding) so that the average bit error probability of the code $\mathcal{C}_m$ tends to zero as $m$ goes to infinity, and $\lim_{m\to\infty} R_m = (1-\varepsilon)C$. From Lemma 3.1, we obtain that $\lim_{m\to\infty} \frac{H(\mathbf{X}_m \mid \mathbf{Y}_m)}{n_m} = 0$ where $\mathbf{X}_m$ and $\mathbf{Y}_m$ designate the transmitted codeword in the code $\mathcal{C}_m$ and the received sequence, respectively, and $n_m$ designates the block length of the code $\mathcal{C}_m$. From Proposition 3.1, we obtain

$$\frac{H(\mathbf{X}_m|\mathbf{Y}_m)}{n_m} \geq 1 - C - (1 - R_m) \cdot$$
$$\cdot \sum_k \left\{ d_{k,m} \sum_{t=0}^{k} \binom{k}{t} (p_1+p_2)^t (p_0+p_3)^{k-t} h_2\left( \frac{1 - \left(1 - \frac{2p_2}{p_1+p_2}\right)^t \left(1 - \frac{2p_3}{p_0+p_3}\right)^{k-t}}{2} \right) \right\}.$$

By letting $m$ go to infinity, then

$$1 - C - \big(1-(1-\varepsilon)C\big) \sum_k \left\{ d_k \sum_{t=0}^{k} \binom{k}{t}(p_1+p_2)^t(p_0+p_3)^{k-t} h_2\left( \frac{1 - \left(\frac{p_1-p_2}{p_1+p_2}\right)^t \left(\frac{p_0-p_3}{p_0+p_3}\right)^{k-t}}{2} \right) \right\} \leq 0$$

and the upper bound on $h_2(\cdot)$ in Lemma 3.2 gives

$$1 - C - \big(1 - (1-\varepsilon)C\big) \cdot$$
$$\cdot \sum_k \left\{ d_k \sum_{t=0}^{k} \binom{k}{t} (p_1+p_2)^t (p_0+p_3)^{k-t} \left[ 1 - \frac{1}{2\ln 2} \left(\frac{p_1-p_2}{p_1+p_2}\right)^{2t} \left(\frac{p_0-p_3}{p_0+p_3}\right)^{2(k-t)} \right] \right\} \leq 0. \quad (24)$$

Since $p_0 + p_1 + p_2 + p_3 = 1$ (i.e., the transition probabilities of the channel in Fig. 1 sum to 1), then

$$\sum_k \left\{ d_k \sum_{t=0}^{k} \binom{k}{t} (p_1+p_2)^t (p_0+p_3)^{k-t} \left[ 1 - \frac{1}{2\ln 2} \left(\frac{p_1-p_2}{p_1+p_2}\right)^{2t} \left(\frac{p_0-p_3}{p_0+p_3}\right)^{2(k-t)} \right] \right\}$$

$$= \sum_k \left\{ d_k \left[ 1 - \frac{1}{2\ln 2} \sum_{t=0}^{k} \binom{k}{t} \left(\frac{(p_1-p_2)^2}{p_1+p_2}\right)^t \left(\frac{(p_0-p_3)^2}{p_0+p_3}\right)^{k-t} \right] \right\}$$

$$= 1 - \frac{1}{2\ln 2} \sum_k \left\{ d_k \sum_{t=0}^{k} \binom{k}{t} \left(\frac{(p_1-p_2)^2}{p_1+p_2}\right)^t \left(\frac{(p_0-p_3)^2}{p_0+p_3}\right)^{k-t} \right\}$$

$$= 1 - \frac{1}{2\ln 2} \sum_k \left\{ d_k \left( \frac{(p_1-p_2)^2}{p_1+p_2} + \frac{(p_0-p_3)^2}{p_0+p_3} \right)^k \right\}$$

$$\leq 1 - \frac{1}{2\ln 2} \left( \frac{(p_1-p_2)^2}{p_1+p_2} + \frac{(p_0-p_3)^2}{p_0+p_3} \right)^{a_{\mathrm{R}}} \quad (25)$$

where $a_{\mathrm{R}} \triangleq \sum_k k d_k$ designates the asymptotic average right degree of the bipartite graphs which refer to the sequence of linear block codes $\{\mathcal{C}_m\}$, and the last transition follows from Jensen's inequality. Combining (24) and (25) gives

$$1 - C - \big(1 - (1-\varepsilon)C\big) \left[ 1 - \frac{1}{2\ln 2} \left( \frac{(p_1-p_2)^2}{p_1+p_2} + \frac{(p_0-p_3)^2}{p_0+p_3} \right)^{a_{\mathrm{R}}} \right] \leq 0.$$

This yields the following lower bound on the asymptotic average right degree:

$$a_{\mathrm{R}} \geq K_1' + K_2' \ln\left(\frac{1}{\varepsilon}\right) \quad (26)$$

where

$$K_1' = -\frac{\ln\left(\frac{1}{2\ln 2}\frac{1-C}{C}\right)}{\ln\left(\frac{(p_1-p_2)^2}{p_1+p_2} + \frac{(p_0-p_3)^2}{p_0+p_3}\right)}, \quad K_2' = -\frac{1}{\ln\left(\frac{(p_1-p_2)^2}{p_1+p_2} + \frac{(p_0-p_3)^2}{p_0+p_3}\right)}. \quad (27)$$

According to Definition 2.2, the density ($\Delta$) of a parity-check matrix is equal to the number of edges in the corresponding bipartite graph normalized per information bit, while the average right degree ($a_R$) is equal to the same number of edges normalized per parity-check node. These different scalings of the number of the edges in a bipartite graph therefore imply that

$$\Delta = \frac{1-R}{R} a_R \quad (28)$$

where $R$ is the rate of a binary linear block code. By our assumption, the asymptotic rate of the sequence of code $\{C_m\}$ is equal to a fraction $1-\varepsilon$ of the capacity. Therefore, by combining (26) and (28) with $R = (1-\varepsilon)C$, we obtain a lower bound on the asymptotic parity-check density which is of the form

$$\frac{K_1 + K_2 \ln\left(\frac{1}{\varepsilon}\right)}{1-\varepsilon}$$

where

$$K_{1,2} = \frac{1-C}{C} \cdot K_{1,2}' \quad (29)$$

and $K_{1,2}'$ are introduced in (27). This completes the proof of the lower bound in (20) with the coefficients $K_{1,2}$ in (21).

*Derivation of the optimization equation* (22): We refer the reader to Appendix A.2, where we also show the existence of such a solution. □

*Discussion:* It is required to show that we achieve an improved lower bound on the parity-check density, as compared to the one in [13, Theorem 2.1]. To this end, it suffices to show that

$$\frac{(p_1-p_2)^2}{p_1+p_2} + \frac{(p_0-p_3)^2}{p_0+p_3} \geq (1-2w)^2. \quad (30)$$

For a proof of this inequality, we refer the reader to Appendix A.3.

This therefore proves that the new lower bound is tighter (i.e., larger) than the original bound in [13, Theorem 2.1] (which corresponds to a two-level quantization of the LLR, as compared to the new bound which is based on a four-level quantization of the LLR).

Based on the proof of Theorem 3.1, we prove and discuss an upper bound on the asymptotic rate of every sequence of binary linear codes for which reliable communication is achievable. The bound refers to soft-decision ML decoding, and it is therefore valid for any suboptimal decoding algorithm. Hence, the following result also provides an upper bound on the achievable rate of ensembles of LDPC codes under iterative decoding where the transmission takes places over an MBIOS channel. The following bound improves the bounds stated in [1, Theorems 1 and 2]:

**Corollary 3.1 ("Four-Level Quantization" Upper Bound on the Asymptotic Achievable Rates of Sequences of Binary Linear Block Codes).** Let $\{C_m\}$ be a sequence of binary linear block codes whose codewords are transmitted with equal probability over an MBIOS channel, and suppose that the block length of this sequences of codes tends to infinity as $m \to \infty$. Let $d_{k,m}$ be the fraction of the parity-check nodes of degree $k$ in an arbitrary representation of the code $C_m$

by a bipartite graph.[2] Then a *necessary condition* for this sequence to achieve *vanishing bit error probability* as $m \to \infty$ is that the asymptotic rate $R$ of this sequence satisfies

$$R \leq 1 - \max \left\{ \frac{1-C}{\sum_k \left\{ d_k \sum_{t=0}^{k} \binom{k}{t} (p_1+p_2)^t (p_0+p_3)^{k-t} h_2 \left( \frac{1 - \left(1 - \frac{2p_2}{p_1+p_2}\right)^t \left(1 - \frac{2p_3}{p_0+p_3}\right)^{k-t}}{2} \right) \right\}}, \right.$$
$$\left. \frac{2(p_2+p_3)}{1 - \sum_k d_k \left(1 - 2(p_2+p_3)\right)^k} \right\} \tag{31}$$

where $p_0, p_1, p_2, p_3$ are introduced in (6), and $d_k$ and $R$ are introduced in (1).

*Proof.* The first term in the RHS of (31) follows from (7) in Proposition 3.1 and (23) in Lemma 3.1. It follows directly by combining both inequalities, and letting the bit error probability $P_\mathrm{b}$ go to zero. The second term in the RHS of (31) follows from the proof of [13, Corollary 3.1] which is based on the erasure decomposition Lemma [11]. □

Considering ensembles of LDPC codes, we note that the fraction $d_k$ of nodes of degree $k$ is calculated in terms of the degree distribution $\rho(\cdot)$ by the equation

$$d_k = \frac{\frac{\rho_k}{k}}{\int_0^1 \rho(x)\,dx} . \tag{32}$$

### 3.2 Extension of the Bounds to $2^d$ Quantization Levels

Following the method introduced in Section 3.1, we commence by deriving a lower bound on the conditional entropy of a transmitted codeword given the received sequence.

**Proposition 3.2.** Let $\mathcal{C}$ be a binary linear block code of length $n$ and rate $R$. Let $\mathbf{x} = (x_1, \ldots x_n)$ and $\mathbf{y} = (y_1, \ldots, y_n)$ designate the transmitted codeword and received sequence, respectively, when the communication takes place over an MBIOS channel with conditional *pdf* $p_{Y|X}(\cdot|\cdot)$. For an arbitrary $d \geq 2$ and $0 \leq l_{2^{d-1}-1} \leq \ldots \leq l_2 \leq l_1 \leq l_0 \triangleq \infty$, let us define the set of probabilities $\{p_s\}_{s=0}^{2^d-1}$ as follows:

$$p_s \triangleq \begin{cases} \Pr\{l_{s+1} < \mathrm{LLR}(Y) \leq l_s \mid X = 0\} & s = 0, \ldots, 2^{d-1} - 2 \\ \Pr\{0 < \mathrm{LLR}(Y) \leq l_{2^{d-1}-1} \mid X = 0\} + \frac{1}{2}\Pr\{\mathrm{LLR}(Y) = 0 \mid X = 0\} & s = 2^{d-1} - 1 \\ \Pr\{-l_{2^{d-1}-1} \leq \mathrm{LLR}(Y) < 0 \mid X = 0\} + \frac{1}{2}\Pr\{\mathrm{LLR}(Y) = 0 \mid X = 0\} & s = 2^{d-1} \\ \Pr\{-l_{2^d-(s+1)} \leq \mathrm{LLR}(Y) < -l_{2^d-s} \mid X = 0\} & s = 2^{d-1}+1, \ldots, 2^d - 1. \end{cases} \tag{33}$$

For an arbitrary parity-check matrix of the code $\mathcal{C}$, let $d_k$ designate the fraction of the parity-checks involving $k$ variables. Then, the conditional entropy of the transmitted codeword given the received

---
[2] For a sequence of ensembles of binary linear codes $\{\mathcal{C}_m\}$, we denote by $d_{k,m}$ the probability of picking (with uniform distribution) a parity-check node of degree $k$ from a bipartite graph which represents the code $\mathcal{C}_m$.

sequence satisfies

$$\frac{H(\mathbf{X}|\mathbf{Y})}{n} \geq 1 - C - (1-R) \sum_k \left\{ d_k \sum_{\substack{k_0,\ldots,k_{2^{d-1}-1} \\ \sum_i k_i = k}} \binom{k}{k_0,\ldots,k_{2^{d-1}-1}} \cdot \right.$$
$$\left. \prod_{i=0}^{2^{d-1}-1} (p_i + p_{2^d-1-i})^{k_i} \, h_2 \left( \frac{1}{2} \left[ 1 - \prod_{i=0}^{2^{d-1}-1} \left( 1 - \frac{2 p_{2^d-1-i}}{p_i + p_{2^d-1-i}} \right)^{k_i} \right] \right) \right\}. \quad (34)$$

*Proof.* Following the proof of Proposition 3.1, we introduce a new physically degraded channel. It is a memoryless binary-input $2^d$-output symmetric channel (see Fig. 1 for $d = 2$). To this end, let $l_1 \geq l_2 \geq \ldots \geq l_{2^{d-1}-1} \in \mathbb{R}^+$ be arbitrary non-negative numbers, and denote $l_0 \triangleq \infty$. The output alphabet of the degraded channel is defined to be $\mathrm{GF}(2^d)$ whose elements form the set

$$\left\{ \sum_{j=0}^{d-1} a_j \, \alpha^j \quad \text{s.t} \quad (a_0, a_1, \ldots, a_{d-1}) \in \{0,1\}^d \right\}.$$

For $s = 0, 1, \ldots, 2^{d-1}-1$, let us denote the $d-1$-bit binary representation of $s$ by $(a_1^{(s)}, a_2^{(s)}, \ldots, a_{d-1}^{(s)})$ i.e.

$$s = \sum_{j=1}^{d-1} a_j^{(s)} \, 2^{j-1}.$$

Let $X_i$ and $Y_i$ designate the random variables referring to the input and output of the original channel $p_{Y|X}(\cdot|\cdot)$ at time $i$ (where $i = 1, 2, \ldots, n$). As a natural generalization of the channel model in Fig. 1, we introduce a channel with $2^d$ quantization levels of the LLR. The output of the degraded channel at time $i$, $z_i$, is calculated from the output $y_i$ of the original channel as follows:

- If $l_{s+1} < \mathrm{LLR}(y_i) \leq l_s$ for some $0 \leq s < 2^{d-1} - 1$, then $z_i = \sum_{j=1}^{d-1} a_j^{(s)} \, \alpha^j$.

- If $0 < \mathrm{LLR}(y_i) \leq l_{2^{d-1}-1}$, then $z_i = \sum_{j=1}^{d-1} \alpha^j$.

- If $-l_{2^{d-1}-1} \leq \mathrm{LLR}(y_i) < 0$, then $z_i = 1 + \sum_{j=1}^{d-1} \alpha^j$.

- If $-l_s \leq \mathrm{LLR}(y_i) < -l_{s+1}$ for some $0 \leq s < 2^{d-1} - 1$, then $z_i = 1 + \sum_{j=1}^{d-1} a_j^{(s)} \, \alpha^j$.

- If $\mathrm{LLR}(y_i) = 0$, then $z_i$ is chosen as $\sum_{j=1}^{d-1} \alpha^j$ or $1 + \sum_{j=1}^{d-1} \alpha^j$ with equal probability ($\frac{1}{2}$).

From (33), the transition probabilities of the degraded channel are expressed in an equivalent way by

$$p_s = \Pr(Z = \sum_{j=1}^{d-1} a_j^{(s)} \, \alpha^j \mid X = 0) = \Pr(Z = 1 + \sum_{j=1}^{d-1} a_j^{(s)} \, \alpha^j \mid X = 1)$$
$$p_{2^d-1-s} = \Pr(Z = 1 + \sum_{j=1}^{d-1} a_j^{(s)} \, \alpha^j \mid X = 0) = \Pr(Z = \sum_{j=1}^{d-1} a_j^{(s)} \, \alpha^j \mid X = 1) \quad (35)$$

where $s = 0, 1, \ldots, 2^{d-1} - 1$. The symmetry in these equalities holds since the channel is MBIOS.

Equations (8)-(12) hold also for the case of $2^d$-level quantization. Thus, we will calculate the entropy of the random variable $Z$, and an upper bound on the entropy of the random vector $\mathbf{Z}$. This will finally provide the lower bound in (34).

Since $X$ is equally likely to be zero or one, the output of the degraded channel, which is symmetric, satisfies the following probability law for $s = 0, 1, \ldots, 2^{d-1} - 1$:

$$\Pr(Z = 1 + \sum_{j=1}^{d-1} a_j^{(s)} \alpha^j) = \Pr(Z = \sum_{j=1}^{d-1} a_j^{(s)} \alpha^j)$$

$$= \Pr(Z = \sum_{j=1}^{d-1} a_j^{(s)} \alpha^j \mid X = 0) \Pr(X = 0) + \Pr(Z = \sum_{j=1}^{d-1} a_j^{(s)} \alpha^j \mid X = 1) \Pr(X = 1)$$

$$= \frac{p_s + p_{2^d - 1 - s}}{2}.$$

The entropy of $Z$ is therefore

$$H(Z) = 2 \sum_{s=0}^{2^{d-1}-1} \frac{p_s + p_{2^d-1-s}}{2} \log_2 \left( \frac{2}{p_s + p_{2^d-1-s}} \right)$$

$$= 1 + \sum_{s=0}^{2^{d-1}-1} (p_s + p_{2^d-1-s}) \log_2 \left( \frac{1}{p_s + p_{2^d-1-s}} \right) \tag{36}$$

where the last transition follows from the equality $\sum_{s=0}^{2^d-1} p_s = 1$.

We now derive an upper bound on the entropy of the random vector $\mathbf{Z}$. To this end, let

$$Z_i = \Theta_i + \sum_{j=1}^{d-1} \Phi_{i,j} \alpha^j, \quad i = 1, 2, \ldots, n. \tag{37}$$

Denoting $\boldsymbol{\Phi}_i = (\Phi_{i,1}, \Phi_{i,2}, \ldots, \Phi_{i,d-1})$, we have that $\boldsymbol{\Theta} = (\Theta_1, \Theta_2, \ldots, \Theta_n)$ is a random vector over $\{0,1\}^n$, and $\boldsymbol{\Phi} = (\boldsymbol{\Phi}_1, \boldsymbol{\Phi}_2, \ldots, \boldsymbol{\Phi}_n)$ is a random vector over $\{0,1\}^{(d-1)n}$. From the decomposition of $Z_i$ in (37), it follows from (35) that

$$\Pr\left(\boldsymbol{\Phi}_i = (a_1^{(s)}, \ldots, a_{d-1}^{(s)})\right) = p_s + p_{2^d-1-s}, \quad i = 1, 2, \ldots, n, \quad s = 0, 1, \ldots, 2^{d-1} - 1. \tag{38}$$

From (37) and (38), and the same chain of equalities leading to (16), it follows that

$$H(\mathbf{Z}) = n \sum_{s=0}^{2^{d-1}-1} (p_s + p_{2^d-1-s}) \log_2 \left( \frac{1}{p_s + p_{2^d-1-s}} \right)$$
$$+ H(\Theta_1, \Theta_2, \ldots, \Theta_n \mid \boldsymbol{\Phi}_1, \boldsymbol{\Phi}_2, \ldots, \boldsymbol{\Phi}_n). \tag{39}$$

As in the proof of Proposition 3.1, we define the syndrome as $\mathbf{S} = \boldsymbol{\Theta} H^T$ where $H$ is a parity-check matrix of the code $\mathcal{C}$. As before, the transmitted codeword $\mathbf{x} \in \mathcal{C}$ only affects the $\boldsymbol{\Theta}$-components of the vector $\mathbf{Z}$ in (37). In parallel to (17), we obtain

$$H(\Theta_1, \Theta_2, \ldots, \Theta_n \mid \boldsymbol{\Phi}_1, \boldsymbol{\Phi}_2, \ldots, \boldsymbol{\Phi}_n) \leq nR + H(\mathbf{S} \mid \boldsymbol{\Phi}_1, \boldsymbol{\Phi}_2, \ldots, \boldsymbol{\Phi}_n). \tag{40}$$

Considering a parity-check equation which involves $k$ variables, let $\{i_1, i_2, \ldots, i_k\}$ be the set of indices of the variables involved in this parity-check equation. The component of the syndrome $\mathbf{S}$ which refers to this parity-check equation is zero if and only if the components of the sub-vector $(\Theta_{i_1}, \Theta_{i_2}, \ldots, \Theta_{i_k})$ differ from the components of the sub-vector $(X_{i_1}, X_{i_2}, \ldots, X_{i_k})$ in an even number of indices. It is clear from (35) that for an index $i$, where $\boldsymbol{\Phi}_i = (a_1^{(s)}, a_2^{(s)}, \ldots, a_{d-1}^{(s)})$ for some $s = 0, \ldots, 2^{d-1} - 1$, the random variables $X_i$ and $\Theta_i$ are different in probability $\frac{p_{2^d-1-s}}{p_s + p_{2^d-1-s}}$.

**Lemma 3.3.** Given that $(\boldsymbol{\Phi}_{i_1}, \boldsymbol{\Phi}_{i_2}, \ldots, \boldsymbol{\Phi}_{i_k})$ has $k_s$ elements of the type $(a_1^{(s)}, a_2^{(s)}, \ldots, a_{d-1}^{(s)})$ (where $s = 0, \ldots, 2^{d-1}-1$), the probability that the components of $(\Theta_{i_1}, \Theta_{i_2}, \ldots, \Theta_{i_k})$ and $(X_{i_1}, X_{i_2}, \ldots, X_{i_k})$ differ in an even number of indices is equal to

$$\frac{1}{2}\left[1 + \prod_{s=0}^{2^{d-1}-1}\left(1 - \frac{2p_{2^d-1-s}}{p_s + p_{2^d-1-s}}\right)^{k_s}\right].$$

*Proof.* The lemma is proved in Appendix B.1. □

Based on Lemma 3.3 and the discussion above, it follows that given that the vector $(\boldsymbol{\Phi}_{i_1}, \boldsymbol{\Phi}_{i_2}, \ldots, \boldsymbol{\Phi}_{i_k})$ has $k_s$ elements of the type $(a_1^{(s)}, a_2^{(s)}, \ldots, a_{d-1}^{(s)})$ then

$$H(S_i \mid \boldsymbol{\Phi}_{i_1}, \boldsymbol{\Phi}_{i_2}, \ldots, \boldsymbol{\Phi}_{i_k}) = h_2\left(\frac{1}{2}\left[1 - \prod_{s=0}^{2^{d-1}-1}\left(1 - \frac{2p_{2^d-1-s}}{p_s + p_{2^d-1-s}}\right)^{k_s}\right]\right).$$

For a component $S_i$ ($1 \leq i \leq n(1-R)$) of the syndrome $\mathbf{S}$ which refers to a parity-check equation involving $k$ variables

$$\begin{aligned}
&H(S_i \mid \boldsymbol{\Phi}_1, \boldsymbol{\Phi}_2, \ldots, \boldsymbol{\Phi}_n) \\
&= H(S_i \mid \boldsymbol{\Phi}_{i_1}, \boldsymbol{\Phi}_{i_2}, \ldots, \boldsymbol{\Phi}_{i_k}) \\
&= \sum_{\underline{\Phi} \in \{0,1\}^{(d-1)k}} \Pr\left((\boldsymbol{\Phi}_{i_1}, \boldsymbol{\Phi}_{i_2}, \ldots, \boldsymbol{\Phi}_{i_k}) = \underline{\Phi}\right) H\left(S_i \mid (\boldsymbol{\Phi}_{i_1}, \boldsymbol{\Phi}_{i_2}, \ldots, \boldsymbol{\Phi}_{i_k}) = \underline{\Phi}\right) \\
&= \sum_{\substack{k_0, \ldots, k_{2^{d-1}-1} \\ \sum_s k_s = k}} \binom{k}{k_0, \ldots, k_{2^{d-1}-1}} \prod_{s=0}^{2^{d-1}-1} (p_s + p_{2^d-1-s})^{k_s} \\
&\quad \cdot h_2\left(\frac{1}{2}\left[1 - \prod_{s=0}^{2^{d-1}-1}\left(1 - \frac{2p_{2^d-1-s}}{p_s + p_{2^d-1-s}}\right)^{k_s}\right]\right).
\end{aligned}$$

The number of parity-check equations involving $k$ variables is $n(1-R)d_k$, hence

$$\begin{aligned}
&H(\mathbf{S} \mid \boldsymbol{\Phi}_1, \boldsymbol{\Phi}_2, \ldots, \boldsymbol{\Phi}_n) \\
&\leq \sum_{i=1}^{n(1-R)} H(S_i \mid \boldsymbol{\Phi}_1, \boldsymbol{\Phi}_2, \ldots, \boldsymbol{\Phi}_n) \\
&= n(1-R) \sum_k \left\{ d_k \sum_{\substack{k_0, \ldots, k_{2^{d-1}-1} \\ \sum_s k_s = k}} \binom{k}{k_0, \ldots, k_{2^{d-1}-1}} \prod_{s=0}^{2^{d-1}-1} (p_s + p_{2^d-1-s})^{k_s} \right. \\
&\quad \left. h_2\left(\frac{1}{2}\left[1 - \prod_{s=0}^{2^{d-1}-1}\left(1 - \frac{2p_{2^d-1-s}}{p_s + p_{2^d-1-s}}\right)^{k_s}\right]\right) \right\}. \qquad (41)
\end{aligned}$$

By combining (39)–(41), an upper bound on the entropy of the random vector $\mathbf{Z}$ follows:

$$H(\mathbf{Z}) \leq nR + n \sum_{s=0}^{2^{d-1}-1} (p_s + p_{2^d-1-s}) \log_2 \left( \frac{1}{p_s + p_{2^d-1-s}} \right)$$

$$+ n(1-R) \sum_k \left\{ d_k \sum_{\substack{k_0,\ldots,k_{2^{d-1}-1} \\ \sum_s k_s = k}} \binom{k}{k_0,\ldots,k_{2^{d-1}-1}} \prod_{s=0}^{2^{d-1}-1} (p_s + p_{2^d-1-s})^{k_s} \right.$$

$$\left. h_2 \left( \frac{1}{2} \left[ 1 - \prod_{s=0}^{2^{d-1}-1} \left( 1 - \frac{2 p_{2^d-1-s}}{p_s + p_{2^d-1-s}} \right)^{k_s} \right] \right) \right\}. \tag{42}$$

The substitution of (36) and (42) in (12) finally provides the lower bound on the conditional entropy $H(\mathbf{X} \mid \mathbf{Y})$ in (34). □

*Discussion:* The calculation of the lower bound in the RHS of (34) becomes more complex as the value of $d$ is increased. However, for *optimally* chosen quantization levels, we show that as the value of $d$ is increased (thus, increasing the number of quantization levels), this bound is monotonically increasing. To this end, let $d \geq 2$ be an arbitrary integer, and let $(l_1^{(d)}, \ldots, l_{2^{d-1}-1}^{(d)})$ and their symmetric values around zero denote the optimal choice for a $2^d$-level quantization. Let $p_0^{(d)}, p_1^{(d)}, \ldots, p_{2^d-1}^{(d)}$ denote the transition probabilities, as defined in (33), which are associated with the optimal $2^d$ quantization levels. In order to show the above monotonicity property, we prove in Appendix B.2 that there exist sub-optimal $2^{d+1}$ quantization levels $\tilde{l}_1, \tilde{l}_2, \ldots, \tilde{l}_{2^d-1}$ (together with their symmetric values around zero) so that even with this sub-optimal $2^{d+1}$-level quantization, the bound in the RHS of (34) is already better than the one which is calculated from the optimal choice of a $2^d$-level quantization.

**Theorem 3.2 ("$2^d$-Level Quantization" Lower Bound on the Asymptotic Parity-Check Density of Binary Linear Block Codes).** Let $\{C_m\}$ be a sequence of binary linear block codes achieving a fraction $1 - \varepsilon$ of the capacity of an MBIOS channel with vanishing *bit error probability*. Then, the asymptotic density of their parity-check matrices satisfies

$$\liminf_{m \to \infty} \Delta_m > \frac{K_1 + K_2 \ln \frac{1}{\varepsilon}}{1 - \varepsilon} \tag{43}$$

where

$$K_1 = K_2 \ln \left( \frac{1}{2 \ln(2)} \frac{1-C}{C} \right), \quad K_2 = -\frac{1-C}{C \ln \left( \sum_{i=0}^{2^{d-1}-1} \frac{(p_i - p_{2^d-1-i})^2}{p_i + p_{2^d-1-i}} \right)}. \tag{44}$$

Here, $d \geq 2$ is an arbitrary integer and the probabilities $\{p_i\}$ are introduced in (33) in terms of $l_1 \geq \ldots \geq l_{2^{d-1}-1} \in \mathbb{R}^+$. The optimal vector of quantization levels $(l_1, \ldots, l_{2^{d-1}-1})$ is given implicitly by the set of $2^{d-1} - 1$ equations

$$\frac{p_{2^d-1-i}^2 + e^{-l_i} p_i^2}{(p_i + p_{2^d-1-i})^2} = \frac{p_{2^d-i}^2 + e^{-l_i} p_{i-1}^2}{(p_{i-1} + p_{2^d-i})^2}, \quad i = 1, \ldots, 2^{d-1} - 1. \tag{45}$$

where such a solution always exists.[3]

---
[3] See the footnote to Theorem 3.1 in p. 10.

*Proof.* For an arbitrary sequence of binary linear block codes $\{\mathcal{C}_m\}$ which achieves a fraction $1-\varepsilon$ to capacity with vanishing bit error probability, we get from Lemma 3.1 that $\lim_{m\to\infty} \frac{H(\mathbf{X}_m \mid \mathbf{Y}_m)}{n_m} = 0$ where $\mathbf{X}_m$ and $\mathbf{Y}_m$ designate the transmitted codeword in the code $\mathcal{C}_m$ and the received sequence, respectively, and $n_m$ designates the block length of the code $\mathcal{C}_m$. From Proposition 3.2, we obtain

$$\frac{H(\mathbf{X}_m|\mathbf{Y}_m)}{n_m} \geq 1 - C - (1 - R_m) \cdot$$

$$\cdot \sum_k \left\{ d_{k,m} \sum_{\substack{k_0,\ldots,k_{2^{d-1}-1} \\ \sum_s k_s = k}} \binom{k}{k_0,\ldots,k_{2^{d-1}-1}} \prod_{s=0}^{2^{d-1}-1} (p_s + p_{2^d-1-s})^{k_s} \right.$$

$$\left. h_2\left( \frac{1}{2}\left[ 1 - \prod_{s=0}^{2^{d-1}-1} \left(1 - \frac{2p_{2^d-1-s}}{p_s + p_{2^d-1-s}}\right)^{k_s} \right] \right) \right\}.$$

Similarly to the proof of Theorem 3.1, by letting $m$ tend to infinity and using the upper bound on $h_2(\cdot)$ from Lemma 3.2, we get

$$1 - C - \left(1 - (1-\varepsilon)C\right) \sum_k \left\{ d_k \sum_{\substack{k_0,\ldots,k_{2^{d-1}-1} \\ \sum_s k_s = k}} \binom{k}{k_0,\ldots,k_{2^{d-1}-1}} \prod_{s=0}^{2^{d-1}-1} (p_s + p_{2^d-1-s})^{k_s} \right.$$

$$\left. \left[ 1 - \frac{1}{2\ln 2}\left( \prod_{s=0}^{2^{d-1}-1} \frac{p_s - p_{2^d-1-s}}{p_s + p_{2^d-1-s}} \right)^{2k_s} \right] \right\} \leq 0. \qquad (46)$$

Since $\sum_k d_k = 1$ and $\sum_{s=0}^{2^d-1} p_s = 1$, the sum in the LHS of (46) is equal to

$$1 - \frac{1}{2\ln 2} \sum_k \left\{ d_k \sum_{\substack{k_0,\ldots,k_{2^{d-1}-1} \\ \sum_s k_s = k}} \binom{k}{k_0,\ldots,k_{2^{d-1}-1}} \prod_{s=0}^{2^{d-1}-1} \left( \frac{(p_s - p_{2^d-1-s})^2}{p_s + p_{2^d-1-s}} \right)^{k_s} \right\}$$

$$= 1 - \frac{1}{2\ln 2} \sum_k \left\{ d_k \left( \sum_{s=0}^{2^{d-1}-1} \frac{(p_s - p_{2^d-1-s})^2}{p_s + p_{2^d-1-s}} \right)^k \right\}$$

$$\leq 1 - \frac{1}{2\ln 2} \left( \sum_{s=0}^{2^{d-1}-1} \frac{(p_s - p_{2^d-1-s})^2}{p_s + p_{2^d-1-s}} \right)^{a_\mathrm{R}} \qquad (47)$$

where $a_\mathrm{R} \triangleq \sum_k k d_k$ designates the asymptotic average right degree, and the last transition follows from Jensen's inequality. Combining (46) and (47) gives

$$1 - C - (1 - (1-\varepsilon)C) \left[ 1 - \frac{1}{2\ln 2} \left( \sum_{s=0}^{2^{d-1}-1} \frac{(p_s - p_{2^d-1-s})^2}{p_s + p_{2^d-1-s}} \right)^{a_\mathrm{R}} \right] \leq 0.$$

This yields the following lower bound on the asymptotic average right degree:

$$a_\mathrm{R} \geq K_1' + K_2' \ln\left(\frac{1}{\varepsilon}\right) \qquad (48)$$

where

$$K_1' = -\frac{\ln\left(\frac{1}{2\ln 2}\frac{1-C}{C}\right)}{\ln\left(\sum_{s=0}^{2^{d-1}-1}\frac{(p_s - p_{2^d-1-s})^2}{p_s + p_{2^d-1-s}}\right)}, \quad K_2' = -\frac{1}{\ln\left(\sum_{s=0}^{2^{d-1}-1}\frac{(p_s - p_{2^d-1-s})^2}{p_s + p_{2^d-1-s}}\right)}.$$

By combining (28) and (48) with the asymptotic rate $R = (1-\varepsilon)C$, we obtain a lower bound on the asymptotic parity-check density which is of the form

$$\frac{K_1 + K_2 \ln\left(\frac{1}{\varepsilon}\right)}{1-\varepsilon}$$

where

$$K_{1,2} = \frac{1-C}{C} \cdot K_{1,2}'.$$

This completes the proof of the lower bound in (43) and (44). The derivation of the set of optimization equations in (45) follows along the lines of the derivation of (22). In the general case of $2^d$ quantization levels, it follows from (44) that we need to maximize

$$\sum_{s=0}^{2^{d-1}-1}\frac{(p_s - p_{2^d-1-s})^2}{p_s + p_{2^d-1-s}}.$$

To this end, we set to zero all the partial derivatives w.r.t. $l_s$ where $s = 1, \ldots, 2^{d-1}-1$. Since from (33) only $p_s$, $p_{s-1}$, $p_{2^d-s}$ and $p_{2^d-s-1}$ depend on $l_s$, then

$$\frac{\partial}{\partial l_s}\left\{\frac{(p_{s-1} - p_{2^d-s})^2}{p_{s-1} + p_{2^d-s}} + \frac{(p_s - p_{2^d-s-1})^2}{p_s + p_{2^d-s-1}}\right\} = 0.$$

We express now the probabilities $p_s$, $p_{s-1}$, $p_{2^d-s}$ and $p_{2^d-s-1}$ as integrals of the conditional *pdf* $a(\cdot)$ of the LLR, and rely on the symmetry property where $a(l) = e^l a(-l)$ for $l \in \mathbb{R}$. In a similar manner to the derivation of (22), this gives the set of equations in (45). Their solution provides the quantization levels $l_1, \ldots, l_{2^{d-1}-1}$ (where according to Proposition 3.2, the other $2^{d-1}-1$ levels are set to be symmetric w.r.t. zero). □

Based on the proof of Theorem 3.2, we derive an upper bound on the asymptotic rate of every sequence of binary linear codes for which reliable communication is achievable. The bound refers of soft-decision ML decoding, and it is therefore valid for any sub-optimal decoding algorithm.

**Corollary 3.2 ("$2^d$-Level Quantization" Upper Bound on the Asymptotic Achievable Rates of Sequences of Binary Linear Block Codes).** Let $\{\mathcal{C}_m\}$ be a sequence of binary linear block codes whose codewords are transmitted with equal probability over an MBIOS channel, and suppose that the block length of this sequences of codes tends to infinity as $m \to \infty$. Let $d_{k,m}$ be the fraction of the parity-check nodes of degree $k$ in an arbitrary representation of the code $\mathcal{C}_m$ by a bipartite graph. Then a *necessary condition* for this sequence to achieve *vanishing bit error probability* as $m \to \infty$ is that the asymptotic rate $R$ of this sequence satisfies

$$R \leq 1 - \max\left\{(1-C)\left\{\sum_k d_k \sum_{\substack{k_0,\ldots,k_{2^{d-1}-1} \\ \cdot \sum_i k_i = k}}\binom{k}{k_0,\ldots,k_{2^{d-1}-1}} \cdot \prod_{i=0}^{2^{d-1}-1}(p_i + p_{2^d-1-i})^{k_i}\right.\right.$$

$$\left.\left. h_2\left(\frac{1}{2}\left[1 - \prod_{i=0}^{2^{d-1}-1}\left(1 - \frac{2p_{2^d-1-i}}{p_i + p_{2^d-1-i}}\right)^{k_i}\right]\right)\right\}^{-1}, \frac{2\sum_{i=2^{d-1}}^{2^d-1} p_i}{1 - \sum_k d_k\left(1 - 2\sum_{i=2^{d-1}}^{2^d-1} p_i\right)^k}\right\} \quad (49)$$

where $d \geq 2$ is arbitrary, the probabilities $\{p_i\}$ are introduced in (33), and $d_k$ and $R$ are introduced in (1).

*Proof.* The concept of the proof is the same as the proof of Corollary 3.1, except that the first term in the RHS of (49) relies on (34). □

## 4 Approach II: Bounds without Quantization of the LLR

Similarly to the previous section, we derive bounds on the asymptotic achievable rate and the asymptotic parity-check density of an arbitrary sequence of binary, linear block codes transmitted over an MBIOS channel. As in Section 3, the derivation of these two bounds is based on a lower bound on the conditional entropy of a transmitted codeword given the received sequence at the output of an arbitrary MBIOS channel.

**Proposition 4.1.** Let $\mathcal{C}$ be a binary linear code of length $n$ and rate $R$ transmitted over an MBIOS channel. Let $\mathbf{x} = (x_1, x_2, \ldots, x_n)$ and $\mathbf{y} = (y_1, y_2, \ldots, y_n)$ be the transmitted codeword and the received sequence, respectively. For an arbitrary representation of the code $\mathcal{C}$ by a parity-check matrix, let $d_k$ designate the fraction of the parity-check equations of degree $k$. Then the conditional entropy of the transmitted codeword given the received sequence satisfies

$$\frac{H(\mathbf{X}|\mathbf{Y})}{n} \geq 1 - C - (1-R)\left(1 - \frac{1}{2\ln(2)} \sum_{p=1}^{\infty} \frac{1}{p(2p-1)} \sum_k d_k \left(\int_0^\infty a(l)(1+e^{-l}) \tanh^{2p}\left(\frac{l}{2}\right) dl\right)^k\right) \quad (50)$$

where $a(\cdot)$ denotes the conditional *pdf* of the LLR given that the transmitted symbol is zero.

*Proof.* We consider a binary linear block code $\mathcal{C}$ of length $n$ and rate $R$ whose transmission takes place over an MBIOS channel. Let us assume that $\mathbf{x} \in \mathcal{C}$ is the transmitted codeword, and the '0' and '1' symbols of the codeword are mapped to $+1$ and $-1$, respectively. The input alphabet to the channel is $\{+1, -1\}$, and the LLR as a function of the observation $y$ at the output of the MBIOS channel gets the form

$$\text{LLR}(y) = \ln\left(\frac{p_{Y|X}(y|X=1)}{p_{Y|X}(y|X=-1)}\right), \quad y \in \mathbb{R}.$$

For the continuation of the proof, we move from the mapping of the MBIOS channel $X \to Y$ to an equivalent representation of the channel $X \to \widetilde{Y}$, so that the conditional entropies of the transmitted codeword given the received sequences at the outputs of both channels are equal, i.e., $H(X|Y) = H(X|\widetilde{Y})$. The basic idea for showing the equivalence between the original channel and the one which will be introduced shortly is based on the following two facts: Firstly, the LLR forms a sufficient statistics of the channel, and secondly, denoting the output of an MBIOS channel by $y$ then, from the symmetry of the channel, $\text{LLR}(-y) = -\text{LLR}(y)$. This means that the absolute value of the LLR doesn't change when the channel output is flipped, but then the sign of the LLR alternates.

For the characterization of the equivalent channel, let $a(\cdot)$ designate the conditional *pdf* of the LLR given that the transmitted symbol is 0 (i.e., given that the channel input is 1). For $i = 1, 2, \ldots, n$, we randomly generate an i.i.d. sequence $\{l_i\}_{i=1}^n$ w.r.t. the conditional *pdf* $a(\cdot)$, and

define

$$\omega_i \triangleq |l_i|, \quad \theta_i \triangleq \begin{cases} +1 & \text{if } l_i > 0 \\ -1 & \text{if } l_i < 0 \\ \pm 1 \text{ w.p. } \frac{1}{2} & \text{if } l_i = 0 \end{cases}.$$

The output of the equivalent channel is defined to be the sequence $\widetilde{\mathbf{y}} = (\widetilde{y}_1, \ldots, \widetilde{y}_n)$ where

$$\widetilde{y}_i = (\phi_i, \omega_i), \quad i = 1, 2, \ldots, n$$

and $\phi_i = \theta_i x_i$. The output of this equivalent channel at time $i$ is therefore the pair $(\phi_i, \omega_i)$ where $\phi_i \in \{+1, -1\}$ and $\omega_i \in \mathbb{R}^+$. This defines the memoryless mapping

$$X \to \widetilde{Y} \triangleq (\Phi, \Omega)$$

where $\Phi$ is a binary random variable which is affected by $X$, and $\Omega$ is a non-negative random variable which represents the absolute value of the LLR and whose *pdf* is

$$f_\Omega(\omega) = \begin{cases} a(\omega) + a(-\omega) = (1 + e^{-\omega}) a(\omega) & \text{if } \omega > 0 \\ a(0) & \text{if } \omega = 0 \end{cases}. \tag{51}$$

We note that the transition in case $\omega > 0$ follows from the symmetry property of $a(\cdot)$, and the random variable $\Omega$ is clearly statistically independent of $X$. Following the lines which lead to (12), we obtain

$$H(\mathbf{X}|\mathbf{Y}) \geq nR - H(\widetilde{\mathbf{Y}}) + nH(\widetilde{Y}) - nC. \tag{52}$$

In order to get a lower bound on $H(\mathbf{X}|\mathbf{Y})$, we will calculate exactly the entropy of $\widetilde{Y}$ and obtain an upper bound on the entropy of $\widetilde{\mathbf{Y}}$. The calculation of the first entropy is direct

$$\begin{aligned} H(\widetilde{Y}) &= H(\Phi, \Omega) \\ &= H(\Omega) + H(\Phi|\Omega) \\ &= H(\Omega) + E_\omega [H(\Phi|\Omega = \omega)] \\ &= H(\Omega) + 1 \end{aligned} \tag{53}$$

where the last transition is due to the fact that given the absolute value of the LLR, its sign is equally likely to be positive or negative. We note that $H(\Omega)$ is not expressed explicitly as it will cancel out later.

We will now derive an upper bound on $H(\widetilde{\mathbf{Y}})$.

$$\begin{aligned} H(\widetilde{\mathbf{Y}}) &= H\big((\Phi_1, \ldots, \Phi_n), (\Omega_1, \ldots, \Omega_n)\big) \\ &= H(\Omega_1, \ldots, \Omega_n) + H\big((\Phi_1, \ldots, \Phi_n), | (\Omega_1, \ldots, \Omega_n)\big) \\ &= nH(\Omega) + H\big((\Phi_1, \ldots, \Phi_n), | (\Omega_1, \ldots, \Omega_n)\big). \end{aligned} \tag{54}$$

Let us introduce the assignment $f : \{+1, -1\} \to \{0, 1\}$ where $+1$ and $-1$ are mapped back to 0 and 1, respectively, and define $\widetilde{\Phi}_i = f(\Phi_i)$. Since $\Phi_i = \Theta_i X_i$, then we obtain

$$\widetilde{\Phi}_i = \widetilde{\Theta}_i + \widetilde{X}_i, \quad i = 1, \ldots, n$$

where the last addition is modulo-2. Define the syndrome vector

$$\mathbf{S} = (\widetilde{\Phi}_1, \ldots, \widetilde{\Phi}_n) H^T$$

where $H$ is an arbitrary parity-check matrix of the binary linear block code $\mathcal{C}$, and let $L$ be the index of the vector $(\widetilde{\Phi}_1, \ldots, \widetilde{\Phi}_n)$ in the coset which corresponds to $\mathbf{S}$. Since each coset has exactly $2^{nR}$ elements which are equally likely then $H(L) = nR$, and we get

$$\begin{aligned}
H\big((\Phi_1,\ldots,\Phi_n), \mid (\Omega_1,\ldots,\Omega_n)\big) &= H(\mathbf{S}, L \mid (\Omega_1,\ldots,\Omega_n)) \\
&\leq H(L) + H(\mathbf{S} \mid (\Omega_1,\ldots,\Omega_n)) \\
&= nR + H(\mathbf{S} \mid (\Omega_1,\ldots,\Omega_n)) \\
&\leq nR + \sum_{j=1}^{n(1-R)} H(S_j \mid (\Omega_1,\ldots,\Omega_n))
\end{aligned} \quad (55)$$

Since $(\widetilde{x}_1, \ldots, \widetilde{x}_n)$ is the original transmitted codeword in $\mathcal{C}$ (i.e., before the conversion of the symbols of the transmitted codeword to $\pm 1$), then

$$\begin{aligned}
\mathbf{S} &= (\widetilde{\Phi}_1, \ldots, \widetilde{\Phi}_n) H^T \\
&= (\widetilde{\Theta}_1, \ldots, \widetilde{\Theta}_n) H^T + (\widetilde{X}_1, \ldots, \widetilde{X}_n) H^T \\
&= (\widetilde{\Theta}_1, \ldots, \widetilde{\Theta}_n) H^T.
\end{aligned}$$

Let us look at the $j$-th parity-check equation which involves $k$ variables. Let us assume that the set of indices of the active variables in this parity-check equation is $\{i_1, \ldots, i_k\}$. Then, the component $S_j$ of the syndrome is equal to 1 if and only if there is an odd number of ones in the random vector $(\widetilde{\Theta}_{i_1}, \ldots, \widetilde{\Theta}_{i_k})$.

**Lemma 4.1.** If the $j$-th component of the syndrome $\mathbf{S}$ involves $k$ active variables in its parity-check equation whose indices are $\{i_1, i_2, \ldots, i_k\}$, then

$$\Pr(S_j = 1 \mid (\Omega_{i_1}, \ldots, \Omega_{i_k}) = (\alpha_1, \ldots, \alpha_k)) = \frac{1}{2}\left[1 - \prod_{m=1}^{k}\left(1 - \frac{2e^{-\alpha_m}}{1 + e^{-\alpha_m}}\right)\right]. \quad (56)$$

*Proof.* The lemma is proved in Appendix C.1. □

We therefore obtain from Lemma 4.1 that

$$H(S_j \mid (\Omega_{i_1}, \ldots, \Omega_{i_k}) = (\alpha_1, \ldots, \alpha_k)) = h_2\left(\frac{1}{2}\left[1 - \prod_{m=1}^{k}\left(1 - \frac{2e^{-\alpha_m}}{1 + e^{-\alpha_m}}\right)\right]\right)$$

and by taking the statistical expectation over the $k$ random variables $\Omega_{i_1}, \ldots, \Omega_{i_k}$, we get

$$\begin{aligned}
&H(S_j \mid (\Omega_{i_1}, \ldots, \Omega_{i_k})) \\
&= \int_0^\infty \cdots \int_0^\infty h_2\left(\frac{1}{2}\left[1 - \prod_{m=1}^{k}\left(1 - \frac{2e^{-\alpha_m}}{1 + e^{-\alpha_m}}\right)\right]\right) \prod_{i=1}^{k} f_\Omega(\alpha_i)\, d\alpha_1 d\alpha_2 \ldots d\alpha_k \quad (57) \\
&= 1 - \frac{1}{2\ln 2} \sum_{p=1}^{\infty} \frac{1}{p(2p-1)} \left(\int_0^\infty f_\Omega(\alpha) \tanh^{2p}\left(\frac{\alpha}{2}\right) d\alpha\right)^k
\end{aligned}$$

where the equality in the last transition is proved in Appendix C.3. Hence, if $d_k$ designates the number of parity-check equations of degree $k$, then

$$\begin{aligned}
&\sum_{j=1}^{n(1-R)} H(S_j \mid (\Omega_1, \ldots, \Omega_n)) \\
&= n(1-R)\left[1 - \frac{1}{2\ln 2} \sum_k d_k \sum_{p=1}^{\infty} \frac{1}{p(2p-1)} \left(\int_0^\infty f_\Omega(\alpha) \tanh^{2p}\left(\frac{\alpha}{2}\right) d\alpha\right)^k\right]. \quad (58)
\end{aligned}$$

By combining (51), (54), (55) and (58), we get the following upper bound on $H(\widetilde{\mathbf{Y}})$:

$$H(\widetilde{\mathbf{Y}}) \leq nH(\Omega) + nR$$
$$+ n(1-R)\left[1 - \frac{1}{2\ln 2}\sum_k d_k \sum_{p=1}^{\infty} \frac{1}{p(2p-1)}\left(\int_0^{\infty} a(\alpha)(1+e^{-\alpha})\tanh^{2p}\left(\frac{\alpha}{2}\right) d\alpha\right)^k\right].\quad(59)$$

Finally, the equality in (53) and the upper bound on $H(\widetilde{\mathbf{Y}})$ given in (59) are substituted in the RHS of (52). This provides the lower bound on the conditional entropy $H(\mathbf{X}|\mathbf{Y})$ given in (50), and completes the proof of this proposition. □

**Remark 4.1.** For the particular case of a BEC with erasure probability $p$, $C = 1 - p$, and the conditional *pdf* of the LLR is independent of the transmitted symbol. It is equal to

$$a(l) = p\Delta_0(l) + (1-p)\Delta_{\infty}(l)$$

where $\Delta_a(\cdot)$ designates the Dirac delta function at the point $a$. We obtain from (50)

$$\frac{H(\mathbf{X}|\mathbf{Y})}{n} \geq p - (1-R)\left[1 - \sum_k d_k(1-p)^k\right]. \quad (60)$$

This lower bound on the conditional entropy for the BEC coincides with the result proved in [13, Eqs. (33) and (34)]. The result there was obtained by the derivation of an upper bound on the rank of $H_{\mathrm{E}}$ which is a sub-matrix of $H$ whose columns correspond to the variables erased by the BEC.

*Discussion:* Since the proof of Proposition 4.1 relies on the analysis of an equivalent channel, rather than a degraded (quantized) channel, it is suggested that the lower bound in the RHS of (50) should be tighter than the one in the RHS of (34). In order to prove this property, it is enough to show that for any integer $d \geq 2$ and any choice of quantization levels $l_1, \ldots, l_{2^{d-1}-1}$, we have

$$1 - \frac{1}{2\ln(2)}\sum_{p=1}^{\infty}\left\{\frac{1}{p(2p-1)}\sum_k d_k \left(\int_0^{\infty} a(l)(1+e^{-l})\tanh^{2p}\left(\frac{l}{2}\right) dl\right)^k\right\}$$
$$\leq \sum_k \left\{d_k \sum_{\substack{k_0,\ldots,k_{2^{d-1}-1} \\ \sum_i k_i = k}} \binom{k}{k_0,\ldots,k_{2^{d-1}-1}}\right.$$
$$\left.\cdot \prod_{i=0}^{2^{d-1}-1}(p_i + p_{2^d-1-i})^{k_i} h_2\left(\frac{1}{2}\left[1 - \prod_{i=0}^{2^{d-1}-1}\left(1 - \frac{2p_{2^d-1-i}}{p_i + p_{2^d-1-i}}\right)^{k_i}\right]\right)\right\} \quad (61)$$

where $p_0, \ldots, p_{2^d-1}$ are the transition probabilities associated with $l_1, \ldots, l_{2^{d-1}-1}$, as defined in (33). This inequality is proved in Appendix D.1, using the power series expansion of the binary entropy function, $h_2(\cdot)$, derived in Appendix C.2.

**Theorem 4.1 ("Un-Quantized" Lower Bound on the Asymptotic Parity-Check Density of Binary Linear Block Codes).** Let $\{C_m\}$ be a sequence of binary linear codes achieving a fraction $1 - \varepsilon$ of the capacity $C$ of an MBIOS channel with vanishing *bit error probability*. Then, the asymptotic density of their parity check matrices satisfies

$$\liminf_{m \to \infty} \Delta_m \geq \sup_{x \in (0,A]} \frac{K_1(x) + K_2(x)\ln\frac{1}{\varepsilon}}{1-\varepsilon} \quad (62)$$

where

$$K_1(x) = \frac{1-C}{C} \frac{\ln\left(\frac{\xi(1-C)}{C}\right)}{\ln\left(\frac{1}{x}\right)}, \quad K_2(x) = \frac{1-C}{C} \frac{1}{\ln\left(\frac{1}{x}\right)}. \tag{63}$$

and

$$A \triangleq \int_0^\infty a(l) \frac{(1-e^{-l})^2}{1+e^{-l}} \, dl, \quad \xi \triangleq \begin{cases} 1 & \text{for a BEC} \\ \frac{1}{2\ln(2)} & \text{otherwise} \end{cases}. \tag{64}$$

*Proof.* From the lower bound on $\frac{H(\mathbf{X} \mid \mathbf{Y})}{n}$ in Eq. (50) and Lemma 3.1 (see p. 10), we obtain that if $\{\mathcal{C}_m\}$ is a sequence of binary linear block codes which achieves a fraction $1-\varepsilon$ of the channel capacity with vanishing bit error probability, then

$$1 - C - \left(1 - (1-\varepsilon)C\right) \left[1 - \frac{1}{2\ln(2)} \sum_k d_k \sum_{p=1}^\infty \frac{1}{p(2p-1)} \left(\int_0^\infty a(l)(1+e^{-l}) \tanh^{2p}\left(\frac{l}{2}\right) dl\right)^k\right] \leq 0. \tag{65}$$

Since $\sum_k k d_k = a_R$ is the average right degree, then from the convexity of the exponential function, we obtain by invoking Jensen's inequality that

$$1 - C - \left(1 - (1-\varepsilon)C\right) \left[1 - \frac{1}{2\ln(2)} \sum_{p=1}^\infty \frac{1}{p(2p-1)} \left(\int_0^\infty a(l)(1+e^{-l}) \tanh^{2p}\left(\frac{l}{2}\right) dl\right)^{a_R}\right] \leq 0. \tag{66}$$

We will now derive two different lower bounds on the infinite sum in the RHS of (66), and compare them later. For the derivation of the lower bound in the first approach, let us define the positive sequence

$$\alpha_p \triangleq \frac{1}{\ln(2)} \frac{1}{2p(2p-1)}, \quad p = 1, 2, \ldots \tag{67}$$

From (C.1) in the appendix, the substitution of $x = 0$ in both sides of the equality gives that $\sum_{p=1}^\infty \alpha_p = 1$, so the sequence $\{\alpha_p\}$ forms a probability distribution. We therefore obtain that

$$\frac{1}{2\ln(2)} \sum_{p=1}^\infty \frac{1}{p(2p-1)} \left(\int_0^\infty a(l)(1+e^{-l}) \tanh^{2p}\left(\frac{l}{2}\right) dl\right)^{a_R}$$

$$= \sum_{p=1}^\infty \alpha_p \left(\int_0^\infty a(l)(1+e^{-l}) \tanh^{2p}\left(\frac{l}{2}\right) dl\right)^{a_R}$$

$$\stackrel{(a)}{\geq} \left(\int_0^\infty a(l)(1+e^{-l}) \sum_{p=1}^\infty \alpha_p \tanh^{2p}\left(\frac{l}{2}\right) dl\right)^{a_R}$$

$$\stackrel{(b)}{=} \left(\int_0^\infty a(l)(1+e^{-l}) \left[1 - h_2\left(\frac{1}{2}\left[1 - \tanh\left(\frac{l}{2}\right)\right]\right)\right] dl\right)^{a_R}$$

$$\stackrel{(c)}{=} \left(\int_0^\infty a(l)(1+e^{-l}) \left[1 - h_2\left(\frac{1}{1+e^{-l}}\right)\right] dl\right)^{a_R}$$

$$\stackrel{(d)}{=} C^{a_R} \tag{68}$$

where inequality (a) follows from Jensen's inequality, equality (b) follows from (67) and (C.1), equality (c) follows from the identity $\tanh(x) = \frac{e^{2x}-1}{e^{2x}+1}$, and equality (d) follows from the relation between the capacity of an MBIOS channel and the *pdf* of the absolute value of the LLR (see [12, Lemma 3.13]).

For an alternative derivation of the lower bound of the infinite series, we can truncate the infinite sum in the RHS of (66) and take into account only the first term in this series. This gives

$$\frac{1}{2\ln(2)} \sum_{p=1}^{\infty} \frac{1}{p(2p-1)} \left( \int_0^{\infty} a(l)(1+e^{-l}) \tanh^{2p}\left(\frac{l}{2}\right) dl \right)^{a_\mathrm{R}}$$

$$\geq \frac{1}{2\ln(2)} \left( \int_0^{\infty} a(l)(1+e^{-l}) \tanh^2\left(\frac{l}{2}\right) dl \right)^{a_\mathrm{R}}$$

$$= \frac{1}{2\ln(2)} \left( \int_0^{\infty} a(l)(1+e^{-l}) \left(\frac{1-e^{-l}}{1+e^{-l}}\right)^2 dl \right)^{a_\mathrm{R}}$$

$$= \frac{A^{a_\mathrm{R}}}{2\ln(2)} \qquad (69)$$

where the last transition follows from (64).

In order to compare the tightness of the two lower bounds in (68) and (69), we first compare the bases of their exponents (i.e., $A$ and $C$). To this end, it is easy to verify that

$$\left(\frac{1-e^{-l}}{1+e^{-l}}\right)^2 \geq 1 - h_2\left(\frac{1}{1+e^{-l}}\right) \quad l \in [0, \infty)$$

with an equality if and only if $l = 0$ or $l \to \infty$. Hence, from (68) and (69), this gives $A \geq C$ with equality if and only if the MBIOS channel is a BEC. Therefore, up to the multiplicative constant $\frac{1}{2\ln(2)}$, the second lower bound is tighter than the first one. However, we note that for the BEC, the first bound is tighter. It gives an improvement by a factor of $2\ln(2) \approx 1.386$.

We will therefore continue the analysis based on the second bound in (69), and then give the potential improvement which follows from the first bound in (68) for a BEC. From (66) and (69), we obtain that

$$1 - C - \left(1 - (1-\varepsilon)C\right)\left(1 - \frac{A^{a_\mathrm{R}}}{2\ln(2)}\right) \leq 0.$$

Hence, one can replace $A$ (where $0 < A \leq 1$) in the last inequality by an arbitrary $x \in (0, A]$, and obtain that the asymptotic average right degree, $a_\mathrm{R}$, satisfies the lower bound

$$a_\mathrm{R} \geq \frac{\ln\left(\frac{1}{2\ln(2)}\left(1 + \frac{1-C}{\varepsilon C}\right)\right)}{\ln\left(\frac{1}{x}\right)}.$$

By dropping the 1 inside the logarithm in the numerator, we obtain that for $x \in (0, A]$

$$a_\mathrm{R} > K_1'(x) + K_2'(x) \ln\left(\frac{1}{\varepsilon}\right) \qquad (70)$$

where $K_1'(x) = \frac{\ln\left(\frac{1}{2\ln(2)}\frac{1-C}{C}\right)}{\ln\left(\frac{1}{x}\right)}$ and $K_2'(x) = \frac{1}{\ln\left(\frac{1}{x}\right)}$. Finally, since the parity-check density and average right degree are related by the equality $\Delta = \left(\frac{1-R}{R}\right) a_\mathrm{R}$, then we obtain the following lower bound on the asymptotic parity-check density:

$$\liminf_{m \to \infty} \Delta_m > \frac{1 - (1-\varepsilon)C}{(1-\varepsilon)C} \left( K_1'(x) + K_2'(x) \ln\left(\frac{1}{\varepsilon}\right) \right)$$

$$> \frac{K_1(x) + K_2(x) \ln\left(\frac{1}{\varepsilon}\right)}{1 - \varepsilon}, \quad \forall\, x \in (0, A] \qquad (71)$$

where $K_{1,2}(x) \triangleq \frac{1-C}{C} K'_{1,2}(x)$. For the BEC, this lower bound can be improved by using the first bound in (68). In this case, $A = C = 1 - p$ where $p$ designates the erasure probability of the BEC, so the additive coefficient $K_1$ in the RHS of (62) is improved to

$$K_1(x) = \frac{p}{1-p} \frac{\ln\left(\frac{p}{1-p}\right)}{\ln\left(\frac{1}{x}\right)}, \quad x \in (0, 1-p].$$

This concludes the proof of this theorem. □

**Remark 4.2.** For a BEC with erasure probability $p$, the maximization of the RHS of (62) yields that the maximal value is achieved for $x = 1 - p$ when $\varepsilon < \frac{p}{1-p}$ (otherwise, if $\varepsilon \geq \frac{p}{1-p}$, the lower bound is useless as it becomes non-positive). Hence, the lower bound on the asymptotic parity-check density stated in Theorem 4.1 coincides with the bound for the BEC in [13, Eq. (3)]. This lower bound was demonstrated in [13, Theorem 2.3] to be tight. This is proved by showing that the sequence of right-regular LDPC ensembles of Shokrollahi [14] is optimal in the sense that it achieves (up to a small additive coefficient) the lower bound on the asymptotic parity-check density for the BEC.

For a general MBIOS channel (other than the BEC), we show in the proof above that the preferable logarithmic growth rate of the lower bound on the parity-check density is achieved by using the bound which follows from (69) (even in the particular case where $x = A$). However, we note that the lower bound on the parity-check density which follows from (68) is *universal* w.r.t. all MBIOS channels with the same capacity.

**Remark 4.3.** The lower bound on the parity-check density in Theorem 4.1 is uniformly tighter than the one in [13, Theorem 2.1] (except for the BSC and BEC where they coincide). For a proof of this claim, the reader is referred to Appendix D.2.

Based on the proof of Theorem 4.1, we prove and discuss an upper bound on the asymptotic rate of every sequence of binary linear codes for which reliable communication is achievable. The bound refers of optimal ML decoding, and is therefore valid for any sub-optimal decoding algorithm. Hence the following result also provides an upper bound on the achievable rate of ensembles of LDPC codes under iterative decoding, where the transmission takes places over an MBIOS channel.

**Corollary 4.1 ("Un-Quantized" Upper Bound on the Asymptotic Achievable Rates of Sequences of Binary Linear Block Codes).** Let $\{\mathcal{C}_m\}$ be a sequence of binary linear block codes whose codewords are transmitted with equal probability over an MBIOS channel, and assume that the block lengths of these codes tend to infinity as $m \to \infty$. Let $d_{k,m}$ be the fraction of the parity-check nodes of degree $k$ for arbitrary representations of the codes $\mathcal{C}_m$ by bipartite graphs. Then a necessary condition on the achievable rate $(R)$ for obtaining vanishing bit error probability as $m \to \infty$ is

$$R \leq 1 - \frac{1-C}{1 - \frac{1}{2\ln(2)} \sum_{p=1}^{\infty} \left\{ \frac{1}{p(2p-1)} \sum_k d_k \left( \int_0^{\infty} a(l) (1 + e^{-l}) \tanh^{2p}\left(\frac{l}{2}\right) dl \right)^k \right\}} \quad (72)$$

where $d_k$ and $R$ are introduced in (1).

*Proof.* This upper bound on the achievable rate follows immediately from Lemma 3.1 (see p. 10) and the lower bound on the conditional entropy in Proposition 4.1. The upper bound on $R$ follows since the bit error probability of the sequence of codes $\{\mathcal{C}_m\}$ vanishes as we let $m$ tend to infinity. □

**Remark 4.4.** We note that the upper bound on the achievable rate in the RHS of (72) doesn't involve maximization, in contrast to the bound in the RHS of (49). The second term of the maximization in the latter bound follows from considerations related to the BEC where such an expression is not required in the RHS of (72). The reader is referred to Appendix D.3 for a proof of this claim.

**Corollary 4.2 (Lower Bounds on the Bit Error Probability of LDPC Codes).** Let $\mathcal{C}$ be a binary linear block code of rate $R$ whose transmission takes place over an MBIOS channel with capacity $C$. For an arbitrary parity-check matrix $H$ of the code $\mathcal{C}$, let $d_k$ designate the fraction of parity-check equations that involve $k$ variables. Then, under ML decoding (or any other decoding algorithm), the bit error probability ($P_\text{b}$) of the code satisfies

$$h_2(P_\text{b}) \geq 1 - \frac{C}{R} + \frac{1-R}{2\ln(2)R} \sum_{p=1}^{\infty} \left\{ \frac{1}{p(2p-1)} \sum_k d_k \left( \int_0^\infty a(l)(1+e^{-l}) \tanh^{2p}\left(\frac{l}{2}\right) dl \right)^k \right\}. \quad (73)$$

*Proof.* This follows directly by combining (23) and (50). □

We now introduce the definition of normalized parity-check density from [13], and derive an improved lower bound on the bit error probability (as compared to [13, Theorem 2.5]) in terms of this quantity.

**Definition 4.1 (Normalized parity-check density [13]).** Let $\mathcal{C}$ be a binary linear code of rate $R$, which is represented by a parity-check matrix $H$ whose density is $\Delta$. The *normalized density* of $H$, call it $t = t(H)$, is defined to be $t = \frac{R\Delta}{2-R}$.

In the following, we clarify the motivation for the definition of a normalized parity-check density. Let us assume that $\mathcal{C}$ is a binary linear block code of length $n$ and rate $R$, and suppose that it can be represented by a bipartite graph which is *cycle-free*. From [13, Lemma 2.1], since this bipartite graph contains $(2-R)n - 1$ edges, connecting $n$ variable nodes with $(1-R)n$ parity-check nodes without any cycles, then the parity-check density of such a cycle-free code is $\Delta = \frac{2-R}{R} - \frac{1}{nR}$. Hence, in the limit where we let $n$ tend to infinity, the normalized parity-check density of a cycle-free code tends to 1. For codes which are represented by bipartite graphs with cycles, the normalized parity-check density is above 1. As shown in [13, Corollary 2.5], the number of fundamental cycles in a bipartite graph which represents an arbitrary linear block $\mathcal{C}$ grows linearly with the normalized parity-check density. The normalized parity-check density therefore provides a measure for the number of cycles in bipartite graphs representing linear block codes. It is well known that cycle-free codes are not good in terms of performance, even under optimal ML decoding [15]; hence, good error-correcting codes (e.g., LDPC codes) should be represented by bipartite graphs with cycles. Following the lead of [13], providing a lower bound on the asymptotic normalized parity-check density in terms of their rate and gap to capacity gives a quantitative measure for the number of fundamental cycles of bipartite graphs representing good error correcting codes. In the following, we provide such an improved bound as compared to the bound given in [13, Theorem 2.5]. In the continuation (see Section 5.2), the resulting improvement is exemplified.

First, we note that from Definition 4.1, it follows that the relation between the normalized parity-check density and the average right degree is

$$t = \left( \frac{1-R}{2-R} \right) a_\text{R}$$

so the normalized parity-check density grows linearly with the average right degree (which is directly linked to the decoding complexity per iteration of LDPC codes under message-passing decoding) where the scaling factor depends on the code rate $R$.

Since $\sum_k k d_k = a_{\mathrm{R}}$, then by applying Jensen's inequality to the RHS of (73), we get the following lower bound on the bit error probability:

$$h_2(P_{\mathrm{b}}) \geq 1 - \frac{C}{R} + \frac{1-R}{2\ln(2)R} \sum_{p=1}^{\infty} \left\{ \frac{1}{p(2p-1)} \left( \int_0^{\infty} a(l)(1+e^{-l}) \tanh^{2p}\left(\frac{l}{2}\right) dl \right)^{\frac{(2-R)t}{1-R}} \right\}. \quad (74)$$

This lower bound on the bit error probability is tighter than the bound given in [13, Eq. (23)] because of two reasons: Firstly, by combining inequality (69) with the inequality proved in Appendix D.2, we obtain that

$$\frac{1}{2\ln(2)} \sum_{p=1}^{\infty} \left\{ \frac{1}{p(2p-1)} \left( \int_0^{\infty} a(l)(1+e^{-l}) \tanh^{2p}\left(\frac{l}{2}\right) dl \right)^{\frac{(2-R)t}{1-R}} \right\} \geq \frac{(1-2w)^{\frac{2(2-R)t}{1-R}}}{2\ln 2}.$$

Secondly, the further improvement in the tightness of the new bound is obtained by dividing the RHS of (74) by $R$ (where $R \leq 1$), as compared to the RHS of [13, Eq. (23)].

The bounds in (73) and (74) become trivial when the RHS of these inequalities are non-positive. Let the (multiplicative) gap to capacity be defined as $\varepsilon \triangleq 1 - \frac{R}{C}$. Analysis shows that the bounds in (73) and (74) are useful unless $\varepsilon \geq \varepsilon_0$ (see Appendices D.4 and D.5). For the bound in the RHS of (73), $\varepsilon_0$ gets the form

$$\varepsilon_0 = \frac{(1-C)B}{C(1-B)}, \quad B \triangleq \frac{1}{2\ln(2)} \sum_{p=1}^{\infty} \left\{ \frac{1}{p(2p-1)} \sum_k d_k \left( \int_0^{\infty} a(l)(1+e^{-l}) \tanh^{2p}\left(\frac{l}{2}\right) dl \right)^k \right\} \quad (75)$$

and for the bound in the RHS of (74), $\varepsilon_0$ is the unique solution of the equation

$$-\varepsilon_0 C + \frac{1-(1-\varepsilon_0)C}{2\ln(2)} \sum_{p=1}^{\infty} \left\{ \frac{1}{p(2p-1)} \left( \int_0^{\infty} a(l)(1+e^{-l}) \tanh^{2p}\left(\frac{l}{2}\right) dl \right)^{\frac{(2-(1-\varepsilon_0)C)t}{1-(1-\varepsilon_0)C}} \right\} = 0. \quad (76)$$

For a proof of (75) and (76), the reader is referred to Appendices D.4 and D.5, respectively. Similarly to [13, Eq. (25)], we note that $\varepsilon_0$ in (76) forms a lower bound on the gap to capacity for an arbitrary sequence of binary linear block codes achieving vanishing bit error probability over an MBIOS channel; the bound is expressed in terms of their asymptotic rate $R$ and normalized parity-check density $t$. It follows from the transition from (73) to (74) that the lower bound on the gap to capacity in (76) is looser as compared to the one given in (75). However, the bound in (76) solely depends on the normalized parity-check density, while the bound in (75) requires full knowledge of the degree distribution for the parity-check nodes.

# 5 Numerical Results

In this section we present numerical results for the information-theoretic bounds on the limitations of binary linear block codes transmitted over MBIOS channels. These results refer to Theorems 3.1, 3.2 and 4.1 and Corollaries 3.1, 3.2, 4.1 and 4.2. As expected, they significantly improve the numerical results presented in [1, Section 4] and [13, Section 4]. This improvement is attributed to the fact that, in contrast to [1, 13], in the derivation of the bounds in this paper, we do not perform a two-level quantization of the LLR which in essence converts the arbitrary MBIOS channel (whose output may be continuous) to a BSC. Throughout this section, we assume transmission of the codes over the binary-input AWGN channel.

## 5.1 Thresholds of LDPC Ensembles under ML Decoding

The following results (see Tables 1–3) provide bounds on the thresholds of LDPC ensembles under ML decoding. They also give an indication on the inherent loss in performance due to the sub-optimality of iterative message-passing decoding.

| LDPC Ensemble | Capacity Limit | 2-Levels Bound [1] | 4-Levels Bound | 8-Levels Bound | Un-Quantized Lower Bound | Upper Bound [5] | DE Threshold |
|---|---|---|---|---|---|---|---|
| (3,6) | +0.187 dB | +0.249 dB | +0.332 dB | +0.361 dB | +0.371 dB | +0.673 dB | +1.110 dB |
| (4,6) | −0.495 dB | −0.488 dB | −0.472 dB | −0.463 dB | −0.463 dB | −0.423 dB | +1.674 dB |
| (3,4) | −0.794 dB | −0.761 dB | −0.713 dB | −0.694 dB | −0.687 dB | −0.510 dB | +1.003 dB |

Table 1: Comparison of thresholds for Gallager's ensembles of regular LDPC codes transmitted over the binary-input AWGN channel. The 2-level lower bound on the threshold of $\frac{E_b}{N_o}$ refers to ML decoding, and is based on [1, Theorem 1] (see also [13, Table II]). The 4-level, 8-level and un-quantized lower bounds apply to ML decoding, and are based on Corollaries 3.1, 3.2 and 4.1, respectively. The upper bound on the threshold of $\frac{E_b}{N_o}$ holds under 'typical pairs' decoding [5] (and hence, also under ML decoding), and the DE thresholds are based on density evolution for iterative message-passing decoding [10].

The upper bounds on the achievable rates derived in [1] and Corollaries 3.1, 3.2 and 4.1 provide lower bounds on the $\frac{E_b}{N_o}$ thresholds under ML decoding. For Gallager's regular LDPC ensembles, the gap between the thresholds under ML decoding and the exact thresholds under the sum-product decoding algorithm (which are calculated using density-evolution analysis) are rather large. For this reason, we also compare the lower bounds on the $\frac{E_b}{N_o}$ thresholds under ML decoding with upper bounds on the $\frac{E_b}{N_o}$ thresholds which rely on "typical pairs decoding" [5]; an upper bound on the $\frac{E_b}{N_o}$ thresholds under an arbitrary sub-optimal decoding algorithm (e.g., "typical pairs decoding") also forms an upper bound on these thresholds under optimal ML decoding. It is shown in Table 1 that the gap between the thresholds under iterative decoding and the bounds for ML decoding (see the columns referring to the DE threshold and the upper bound based on "typical pairs decoding") is rather large. This is attributed to the sub-optimality of belief propagation decoding for regular LDPC ensembles. On the other hand, it is also demonstrated in Table 1 that the gap between the upper and lower bounds on the thresholds under ML decoding is much smaller. For example, according to the numerical results in Table 1, the inherent loss in the asymptotic performance due to the sub-optimality of belief propagation for Gallager's ensemble of (4, 6) regular LDPC codes

(whose design rate is $\frac{1}{3}$ bits per channel use) ranges between 2.097 and 2.137 dB.

For irregular LDPC ensembles, the calculation of similar upper bounds based on "typical pairs" decoding [5] is based on the calculation of the asymptotic growth rate of the distance spectra of such ensembles. For a given pair of degree distributions $(\lambda, \rho)$, the calculation of the asymptotic growth rates of the distance spectrum for the $(n, \lambda, \rho)$ LDPC ensemble (where we let $n$ tend to infinity) is tractable (see [2, 17]). However, we avoid calculating these upper bounds in Tables 2 and 3 due to the fact that the gap between the DE thresholds under belief propagation and the improved lower bounds on the $\frac{E_b}{N_o}$ thresholds derived in this paper is already rather small (see Tables 2 and 3). The rather small gap between the DE thresholds and the un-quantized lower bounds on the thresholds under ML decoding also indicate that for the degree distributions which are provided by the LDPC optimizer [16], the asymptotic degradation in performance due to the sub-optimality of belief propagation is marginal (it is observed from Tables 2 and 3 that for several LDPC ensembles, this degradation in the asymptotic performance is at most in the order of hundredthes of a decibel).

| $\lambda(x)$ | $\rho(x)$ | 2-Levels Bound [1] | 4-Levels Bound | 8-Levels Bound | Un-Quantized Lower Bound | DE Threshold |
|---|---|---|---|---|---|---|
| $0.38354x + 0.04237x^2 + 0.57409x^3$ | $0.24123x^4 + 0.75877x^5$ | 0.269 dB | 0.370 dB | 0.404 dB | 0.417 dB | 0.809 dB |
| $0.23802x + 0.20997x^2 + 0.03492x^3 + 0.12015x^4 + 0.01587x^6 + 0.00480x^{13} + 0.37627x^{14}$ | $0.98013x^7 + 0.01987x^8$ | 0.201 dB | 0.226 dB | 0.236 dB | 0.239 dB | 0.335 dB |
| $0.21991x + 0.23328x^2 + 0.02058x^3 + 0.08543x^5 + 0.06540x^6 + 0.04767x^7 + 0.01912x^8 + 0.08064x^{18} + 0.22798x^{19}$ | $0.64854x^7 + 0.34747x^8 + 0.00399x^9$ | 0.198 dB | 0.221 dB | 0.229 dB | 0.232 dB | 0.310 dB |
| $0.19606x + 0.24039x^2 + 0.00228x^5 + 0.05516x^6 + 0.16602x^7 + 0.04088x^8 + 0.01064x^9 + 0.00221x^{27} + 0.28636x^{29}$ | $0.00749x^7 + 0.99101x^8 + 0.00150x^9$ | 0.194 dB | 0.208 dB | 0.214 dB | 0.216 dB | 0.274 dB |

Table 2: Comparison of thresholds for rate one-half ensembles of irregular LDPC codes transmitted over the binary-input AWGN channel. The Shannon capacity limit corresponds to $\frac{E_b}{N_o} = 0.187$ dB. The 2-level, 4-level, 8-level and un-quantized lower bounds on the threshold refer to ML decoding, and are based on [1, Theorem 2], Corollaries 3.1, 3.2 and 4.1, respectively. The degree distributions of the ensembles and their DE thresholds are based on density evolution under iterative message-passing decoding [10], and are taken from [11, Tables 1 and 2].

The plots in Figure 2 compare different lower bounds on the $\frac{E_b}{N_0}$-threshold under ML decoding of right-regular LDPC ensembles. The plots refer to a right degree of 6 (left plot) or 10 (right plot). The following lower bounds are depicted in these plots: the Shannon capacity limit, the 2-level quantization lower bound in [1, Theorem 1], the 4 and 8-level quantization bounds of the LLR in Section 3, and finally, the bound in Section 4 where no quantization of the LLR is performed. It can be observed from the two plots in Figure 2 that the range of code rates where there exists a visible improvement with the new lower bounds depends on the degree of the parity-check nodes. In principle, the larger the value of the right-degree is, then the improvement obtained by these bounds is more pronounced starting from a higher rate code rate (e.g., for a right degree of 6 or

| $\lambda(x)$ | $\rho(x)$ | 2-Levels Bound [1] | 4-Levels Bound | 8-Levels Bound | Un-Quantized Lower Bound | DE Threshold |
|---|---|---|---|---|---|---|
| $0.302468x + 0.319447x^2 + 0.378085x^4$ | $x^{11}$ | 1.698 dB | 1.786 dB | 1.815 dB | 1.825 dB | 2.049 dB |
| $0.244067x + 0.292375x^2 + 0.463558x^6$ | $x^{13}$ | 1.664 dB | 1.718 dB | 1.736 dB | 1.742 dB | 1.874 dB |
| $0.205439x + 0.255432x^2 + 0.0751187x^4 + 0.1013440x^5 + 0.3626670x^{11}$ | $x^{15}$ | 1.647 dB | 1.680 dB | 1.691 dB | 1.695 dB | 1.763 dB |

Table 3: Comparison of thresholds for rate-$\frac{3}{4}$ ensembles of irregular LDPC codes transmitted over the binary-input AWGN channel. The Shannon capacity limit corresponds to $\frac{E_b}{N_o} = 1.626$ dB. The 2-level, 4-level, 8-level and un-quantized lower bounds on the threshold refer to ML decoding, and are based on [1, Theorem 2], Corollaries 3.1, 3.2 and 4.1, respectively. The degree distributions of the ensembles and their DE thresholds are based on density evolution under iterative message-passing decoding [10], and are taken from [16].

10, the improvement obtained by the new bounds is observed for code rates starting from 0.35 and 0.55 bits per channel use, respectively).

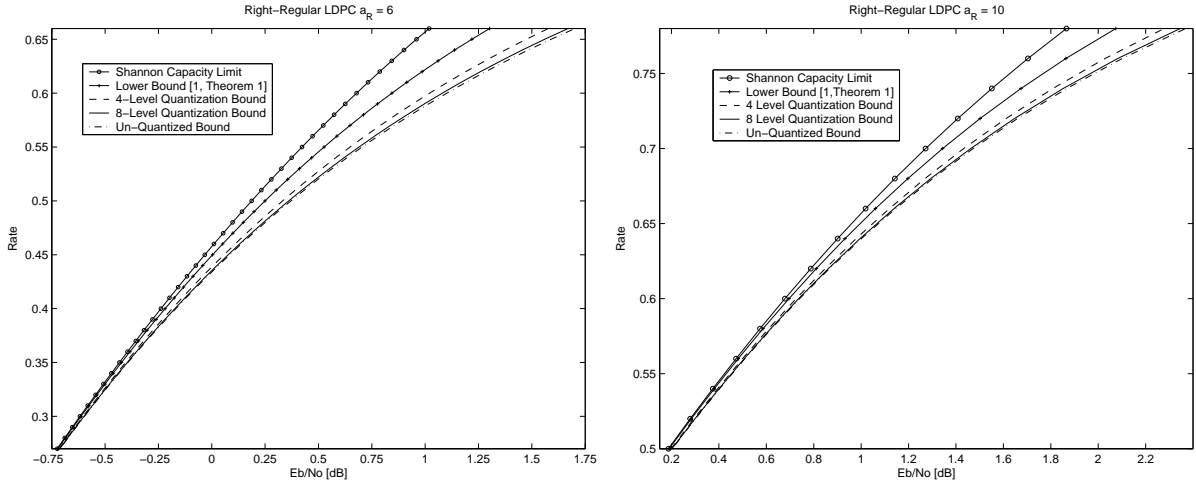

Figure 2: Comparison between different lower bounds on the threshold under ML decoding for right-regular LDPC ensembles with $a_R = 6$ (left plot) and $a_R = 10$ (right plot). The transmission takes place over the binary-input AWGN channel.

## 5.2 Lower Bounds on the Bit Error Probability of LDPC Codes

By combining the lower bound in Proposition 4.1 and Lemma 3.1, we obtain in Corollary 4.2 an improved lower bound on the bit error probability of binary linear block codes, as compared to the one given in [13, Theorem 2.5]. The plot of Fig. 3 presents a comparison of these lower bounds for binary linear block codes where the bounds rely on (74) and [13, Theorem 2.5]. They are plotted as a function of the normalized density of an arbitrary parity-check matrix. In our setting, the capacity of the channel is $\frac{1}{2}$ bit per channel use, and the bounds are depicted for binary linear block codes whose rate is a fraction $1 - \varepsilon$ of the channel capacity. To demonstrate the advantage of the lower bound on the bit error probability in (74) over the lower bound derived in [13, Theorem 2.5], let us assume that one wants to design a binary LDPC code which achieves a bit-error probability of $10^{-6}$ at a rate which is 99% of the channel capacity. The curve of the lower bound from [13] for $\varepsilon = 0.01$ implies that the normalized density of an arbitrary parity-check matrix which represents the code (see Definition 4.1 in p. 27) should be at least 4.33, while the curve depicting the bound from (74) strengthens this requirement to a normalized density (of each parity-check matrix) of at least 5.68. Translating this into terms of parity-check density (which is also the complexity per iteration for message-passing decoding) yields minimal parity-check densities of 13.16 and 17.27, respectively (the minimal parity-check density is given by $\Delta_{\min} = \frac{(2-R)t_{\min}}{R}$). It is reflected from Fig. 3 that as the gap to capacity $\varepsilon$ tends to zero, the lower bound on the normalized density of an arbitrary parity-check matrix ($t$), which represents a code which achieves low error probability for a rate of $R = (1-\varepsilon)C$ grows significantly.

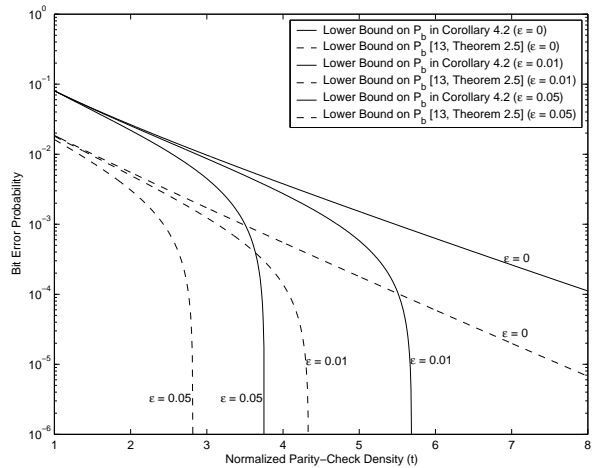

Figure 3: Lower bounds on the bit error probability for any binary linear block code transmitted over a binary-input AWGN channel whose capacity is $\frac{1}{2}$ bits per channel use. The bounds are depicted in terms of the normalized density of an arbitrary parity-check matrix which represents the code, and the curves correspond to code rates which are a fraction $1-\varepsilon$ of the channel capacity (for different values of $\varepsilon$). The bounds depicted in dashed lines are based on [13, Theorem 2.5], and the bounds in solid lines are given in Corollary 4.2.

### 5.3 Lower Bounds on the Asymptotic Parity-Check Density

The lower bound on the parity-check density derived in Theorem 4.1 enables to assess the tradeoff between asymptotic performance and asymptotic decoding complexity (per iteration) of an iterative message-passing decoder. This bound tightens the lower bound on the asymptotic parity-check density derived in [13, Theorem 2.1]. Fig. 4 compares these bounds for codes of rate $\frac{1}{2}$ (left plot) and $\frac{3}{4}$ (right plot) where the bounds are plotted as a function of $\frac{E_b}{N_0}$. It can be observed from Fig. 4 that as $\frac{E_b}{N_0}$ increases, the advantage of the bound in Theorem 4.1 over the bound in [13, Theorem 2.1] diminishes. This follows from the fact that as the value of $\frac{E_b}{N_0}$ is increased, the two-level quantization of the LLR used in [1] and [13, Theorem 2.1] better captures the true behavior of the MBIOS channel. It is also reflected in this figure that as $\varepsilon$ tends to zero (i.e., when the gap to capacity vanishes), the slope of the bounds becomes very sharp. This is due to the logarithmic behavior of the bounds.

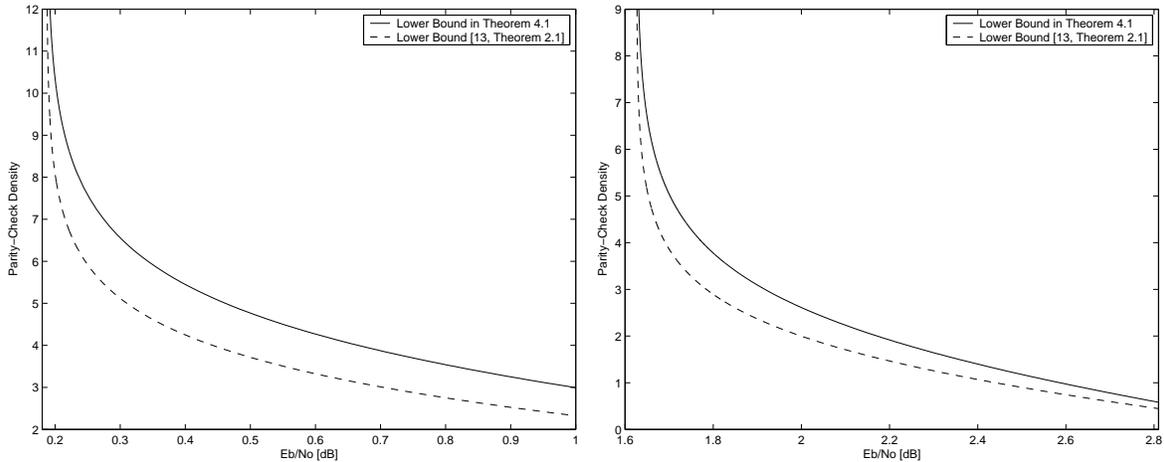

Figure 4: Comparison between lower bounds on the asymptotic parity-check density of binary linear block codes where the transmission takes place over a binary-input AWGN channel. The dashed line refers to [13, Theorem 2.1], and the solid line refers to Theorem 4.1. The left and right plots refer to code rates of $\frac{1}{2}$ and $\frac{3}{4}$, respectively. The Shannon capacity limit for these code rates corresponds to $\frac{E_b}{N_0}$ of 0.187 dB and 1.626 dB, respectively.

# 6  Summary and Outlook

We derive improved lower bounds on the asymptotic density of parity-check matrices and upper bounds on the achievable rates of binary linear block codes transmitted over memoryless binary-input output-symmetric (MBIOS) channels. The improvements are w.r.t. the bounds given in [1, 13]. The information-theoretic bounds are valid for *every* sequence of binary linear block codes, in contrast to high probability results which follow from probabilistic analytical tools (e.g., density evolution (DE) analysis under iterative decoding). The bounds hold under optimal ML decoding, and hence, they hold in particular under any sub-optimal decoding algorithm. We apply the bounds to ensembles of low-density parity-check (LDPC) codes. The significance of the bounds is the following: firstly, by comparing the new upper bounds on the achievable rates with thresholds provided by DE analysis, we obtain rigorous bounds on the inherent loss in performance of various LDPC ensembles. This degradation in the asymptotic performance is due to the sub-optimality of iterative message-passing decoding (as compared to optimal ML decoding). Secondly, the parity-check density can be interpreted as the complexity per iteration under iterative message-passing decoding. Therefore, by tightening the reported lower bound on the asymptotic parity-check density [13, Theorem 2.1], the new bounds provide better insight on the tradeoff between the asymptotic performance and the asymptotic decoding complexity of iteratively decoded LDPC codes. Thirdly, the new lower bound on the bit error probability of binary linear block codes tightens the reported lower bound in [13, Theorem 2.5].

The derivation of the bounds in Section 3 was motivated by the desire to generalize the results in [1, Theorems 1 and 2] and [13, Theorem 2.1]. The two-level quantization of the log-likelihood ratio (LLR) which in essence replaces the arbitrary MBIOS channel by a physically degraded binary symmetric channel (BSC), is modified in Section 3 to a quantized channel which better reflects the statistics of the original channel (though the quantized channel is still physically degraded w.r.t. the original channel). The number of quantization levels at the output of the new channel is an arbitrary integer power of 2. The calculation of the bounds in Section 3 is subject to an optimization of the quantization levels of the LLR, as to get the tightest bounds within their form.

In Section 4, we rely on the conditional *pdf* of the LLR at the output of the MBIOS channel, and operate on an equivalent channel without quantizing the LLR. This second approach finally leads to bounds which are uniformly tighter than the bounds we derive in Section 3. It appears to be even simpler to calculate the un-quantized bounds in Section 4, as their calculation do not involve the solution of any optimization equation (in contrast to the quantized bounds, whose calculation involves a numerical solution of optimization equations w.r.t. the quantization levels of the LLR). The comparison between the quantized and un-quantized bounds gives insight on the effect of the number of quantization levels of the LLR (even if they are chosen optimally) on the achievable rates, as compared to the ideal case where no quantization is done. The results of such a comparison are shown in Tables 1–3, and indicate that the improvement in the tightness of the bounds when more than 8 levels of quantization are used (in case the quantization levels are optimally determined) is marginal. We also note that practically, the possibility to calculate un-quantized bounds which are uniformly better than the quantized bounds was facilitated due to an efficient transformation of the multi-dimensional integral in Appendix C.3 into an infinite series of one-dimensional integrals whose convergence rate is very fast (10 terms are sufficient in practice, as justified in Appendix C.3). Had we used instead an upper bound on $h_2(\cdot)$ like the one in Lemma 3.2 (see p. 10),[4] then this would loosen the un-quantized bounds, and make them sometimes even worse than the 8-level quantized bounds. The ability to express the $k$-dimensional integral in Appendix C.3 in a closed form was therefore crucial for practically obtaining an un-quantized bound which is uniformly tighter than the quantized bounds for an arbitrary number of quantization levels. As mentioned before, the exact calculation of the un-quantized bound also provides insight on the effect of the number of quantization levels on the tightness of the quantized bounds (see Tables 1–3 in Section 5).

Our bounds on the thresholds of LDPC ensembles under optimal ML decoding depend only on the degree distribution of their parity-check nodes and their design rate. For a given parity-check degree distribution ($\rho$) and design rate ($R$), the bounds provide an indication on the inherent gap to capacity which is independent of the choice of $\lambda$ (as long as the pair of degree distributions ($\lambda, \rho$) yield the design rate $R$). Therefore, our bounds are not expected to be tight for LDPC ensembles with a given pair of degree distributions ($\lambda, \rho$). The numerical results shown in Tables 1–3 indicate, however, that these bounds are useful for assessing the inherent gap to capacity of various (regular and irregular) LDPC ensembles. The comparison of our bounds with the DE thresholds (based on density evolution) provides an assessment of the degradation in the asymptotic performance due to the sub-optimality of the iterative sum-product decoding algorithm. We note that the gap of LDPC ensembles to capacity is an inherent phenomenon, due to their finite average-right degree [13]. As a topic for further research, it is suggested to examine the possibility of tightening the bounds for specific ensembles by explicitly taking into account the exact characterization of $\lambda$. We also suggest to study a possible generalization of the bounds to non-binary linear block codes. These generalized bounds can be applied to the analysis of the ML performance of non-binary LDPC ensembles whose transmission takes place over arbitrary discrete memoryless channels with possibly different types of quantization [3].

The lower bound on the asymptotic parity-check density in [13, Theorem 2.1] and its improvements in Sections 3 and 4 grow like the log of the inverse of the gap (in rate) to capacity. The result in [13, Theorem 2.3] shows that a logarithmic growth rate of the parity-check density is achievable for Gallager's regular LDPC ensemble under ML decoding when transmission takes place over an arbitrary MBIOS channel. These results show that for any iterative decoder which is based on the representation of the codes by Tanner graphs, there exists a tradeoff between asymptotic performance and complexity which cannot be surpassed. Recently, it was shown in [9] that better tradeoffs can be achieved by allowing more complicated graphical models; for the particular case of the binary erasure channel (BEC), the encoding and the decoding complexity of properly designed codes on graphs remain bounded as the gap to capacity vanishes. To this end, Pfister, Sason and Urbanke consider in [9, Theorems 1 and 2] ensembles of irregular repeat-accumulate codes which involve punctured bits, and allow in this way a sufficient number of state nodes in the Tanner graph representing the codes. This surprising result is considered in [9, Theorem 4], by a derivation of an information-theoretic lower bound on the decoding complexity of randomly punctured codes on graphs whose transmission takes place over MBIOS channels. The approach for the derivation of the bounds in [9, Theorem 4] rely on the analysis in [13, Theorem 2.1]. As a topic for further research, we suggest to tighten the lower bounds in [9, Theorem 4] by relying on the approach used to prove the improved lower bound on the parity-check density stated in Theorem 4.1 (as compared to the derivation of the bound in [9, Theorem 4] which relies on the analysis in [13, Theorem 2.1]).

---

[4]This was actually the way we used to calculate the un-quantized bounds at the beginning, due to an initial difficulty in calculating the $k$ dimensional integral in Appendix C.3.

# Appendix A

## A.1 Proof of Lemma 3.1

In order to prove Lemma 3.1, let $P_{\text{b}}^{(i)}$ designate the bit error probability of the $i$-th symbol in the code $\mathcal{C}$ (where $1 \leq i \leq n$). Therefore, $P_{\text{b}} = \frac{1}{nR} \sum_{i=1}^{nR} P_{\text{b},i}$ is the average bit error probability of the code, and

$$
\begin{aligned}
\frac{H(\mathbf{X} \mid \mathbf{Y})}{n} &\stackrel{(a)}{=} \frac{\sum_{i=1}^{n} H(x_i \mid \mathbf{Y}, x_1, x_2, \ldots, x_{i-1})}{n} \\
&\stackrel{(b)}{\leq} \frac{\sum_{i=1}^{nR} H(x_i \mid \mathbf{Y})}{n} + \frac{\sum_{i=nR+1}^{n} H(x_i \mid x_1, x_2, \ldots, x_{nR})}{n} \\
&\stackrel{(c)}{=} \frac{\sum_{i=1}^{nR} H(x_i \mid \mathbf{Y})}{n} \\
&\stackrel{(d)}{\leq} \frac{\sum_{i=1}^{nR} h_2(P_{\text{b},i})}{n} \\
&\stackrel{(e)}{\leq} R\, h_2\left(\frac{1}{nR} \sum_{i=1}^{nR} P_{\text{b},i}\right) \\
&= R\, h_2(P_{\text{b}})
\end{aligned}
$$

where equality (a) follows from the chain rule of the entropy, inequality (b) is due to the fact that conditioning reduces the entropy, equality (c) follows since the dimension of the code is $nR$ which implies that there is a set of $nR$ bits of the codeword whose knowledge reveals the entire codeword (w.l.g., one can assume that these are the first $nR$ bits), inequality (d) is based on Fano's inequality and since the code is binary, and inequality (e) is based on Jensen's inequality and the concavity of the binary entropy function.

## A.2 Derivation of the Optimization Equation in (22) and Proving the Existence of its Solution

*Derivation of the optimization equation* (22): We derive here the optimization equation (22) which refers to the "four-level quantization" lower bound on the parity-check density (see p. 10).

Let $a(\cdot)$ designate the conditional *pdf* of the LLR at the output of the original MBIOS channel, given the zero symbol is transmitted. In the following, we express the transition probabilities of the degraded channel in Fig. 1 (see p. 6) in terms of $a(\cdot)$ and the value of $l$:

$$p_0 = \Pr(Z = 0 \mid X = 0) = \int_l^\infty a(u)\, du \tag{A.1}$$

$$p_1 = \Pr(Z = \alpha \mid X = 0) = \int_{0^+}^l a(u)\, du + \frac{1}{2}\int_{0^-}^{0^+} a(u)\, du \tag{A.2}$$

$$p_2 = \Pr(Z = 1 + \alpha \mid X = 0) = \int_{-l}^{0^-} a(u)\, du + \frac{1}{2}\int_{0^-}^{0^+} a(u)\, du \tag{A.3}$$

$$p_3 = \Pr(Z = 1 \mid X = 0) = \int_{-\infty}^{-l} a(u)\, du. \tag{A.4}$$

We note that the integration of $a(\cdot)$ from $u = 0^-$ to $u = 0^+$ is meaningful if and only if there is a non-vanishing probability that the value of the LLR at the output of the original channel is zero (e.g., a BEC). Otherwise, the contribution of this integral to (A.2) and (A.3) vanishes. Since the channel is MBIOS, the symmetry property [10] gives

$$a(u) = e^u\, a(-u), \quad \forall\, u \in \mathbb{R}. \tag{A.5}$$

Based on the expressions for the coefficients $K_1$ and $K_2$ in the lower bound on the asymptotic parity-check density (20), then in order to find the tightest lower bound then we need to maximize

$$\frac{(p_1 - p_2)^2}{p_1 + p_2} + \frac{(p_0 - p_3)^2}{p_0 + p_3} \tag{A.6}$$

w.r.t. the free parameter $l \in \mathbb{R}^+$. From Eqs. (A.1)–(A.4) and the symmetry property in (A.5)

$$p_0 - p_3 = \int_l^\infty a(u)(1 - e^{-u})\, du \Rightarrow \frac{\partial}{\partial l}(p_0 - p_3) = -a(l)(1 - e^{-l}) \tag{A.7}$$

$$p_0 + p_3 = \int_l^\infty a(u)(1 + e^{-u})\, du \Rightarrow \frac{\partial}{\partial l}(p_0 + p_3) = -a(l)(1 + e^{-l}) \tag{A.8}$$

$$p_1 - p_2 = \int_{0^+}^l a(u)(1 - e^{-u})\, du \Rightarrow \frac{\partial}{\partial l}(p_1 - p_2) = a(l)(1 - e^{-l}) \tag{A.9}$$

$$p_1 + p_2 = \int_{0^+}^l a(u)(1 + e^{-u})\, du \Rightarrow \frac{\partial}{\partial l}(p_1 + p_2) = a(l)(1 + e^{-l}) \tag{A.10}$$

so the calculation of the partial derivative of (A.6) w.r.t. $l$ gives

$$\frac{\partial}{\partial l}\left\{\frac{(p_1 - p_2)^2}{p_1 + p_2} + \frac{(p_0 - p_3)^2}{p_0 + p_3}\right\}$$
$$= -4\, a(l)\left\{\left[\left(\frac{p_2}{p_1 + p_2}\right)^2 - \left(\frac{p_3}{p_0 + p_3}\right)^2\right] + e^{-l}\left[\left(\frac{p_1}{p_1 + p_2}\right)^2 - \left(\frac{p_0}{p_0 + p_3}\right)^2\right]\right\}.$$

Since the first derivative of a function changes its sign at a neighborhood of any local maxima or minima point, and since $a(\cdot)$ is always non-negative, then the second multiplicative term above is the one which changes its sign at a neighborhood of $l$ maximizing (A.6). For this value of $l$, the second multiplicative term vanishes, which gives the optimization equation for $l$ in (22).

*Proof of existence of a solution to* (22): In order to show that a solution to (22) always exists, we will see how the LHS and the RHS of this equation behave as $l \to 0^+$ and $l \to \infty$. From (A.1)–(A.4), it follows that in the limit where $l \to \infty$, we get

$$p_1 \to 1 - w - \Pr(\text{LLR}(Y) = \infty \mid X = 0), \quad p_2 \to w$$

where $w$ is introduced in (2), and therefore

$$\lim_{l \to \infty} \frac{p_2^2 + e^{-l} p_1^2}{(p_1 + p_2)^2} = \left( \frac{w}{1 - \Pr(\text{LLR}(Y) = \infty \mid X = 0)} \right)^2. \tag{A.11}$$

Since from the symmetry property

$$p_3 = \int_l^\infty a(-u) du = \int_l^\infty e^{-u} a(u) du \leq e^{-l} \int_l^\infty a(u) du = e^{-l} p_0$$

then the fraction $\frac{p_3}{p_0}$ tends to zero as $l \to \infty$, so

$$\lim_{l \to \infty} \frac{p_3^2 + e^{-l} p_0^2}{(p_0 + p_3)^2} = \lim_{l \to \infty} \frac{\left(\frac{p_3}{p_0}\right)^2 + e^{-l}}{\left(1 + \frac{p_3}{p_0}\right)^2} = 0. \tag{A.12}$$

It therefore follows from (A.11) and (A.12) that for large enough values of $l$, the LHS of (22) is larger than the RHS of this equation. On the other hand, in the limit where $l \to 0^+$, we get

$$p_1, p_2 \to \frac{1}{2} \int_{0^-}^{0^+} a(u) du$$

and therefore

$$\lim_{l \to 0^+} \frac{p_2^2 + e^{-l} p_1^2}{(p_1 + p_2)^2} = \frac{1}{2}. \tag{A.13}$$

In the limit where $l \to 0^+$

$$p_0 \to \int_{0^+}^\infty a(u) du, \quad p_3 \to \int_{-\infty}^{0^-} a(u) du, \quad p_0 + p_3 \to \beta$$

where $\beta \triangleq 1 - \int_{0^-}^{0^+} a(u) du$. By denoting $u \triangleq \int_{0^+}^\infty a(u) du$, we get $0 \leq u \leq \beta$, and

$$\lim_{l \to 0^+} \frac{p_3^2 + e^{-l} p_0^2}{(p_0 + p_3)^2} = \frac{u^2 + (\beta - u)^2}{\beta^2} \geq \frac{1}{2}, \quad \forall u \in [0, \beta]. \tag{A.14}$$

We note that the last inequality holds in equality if and only if $u = \frac{\beta}{2}$. But if this condition holds, then this implies that

$$\int_{-\infty}^{0^-} a(u) \, du = \int_{0^+}^\infty a(u) \, du$$

which from the symmetry property cannot be satisfied unless $a(u) = \delta(u)$. The latter condition corresponds to a BEC with erasure probability 1 (whose capacity is equal to zero).

From (A.13) and (A.14), we obtain that for small enough (and non-negative) values of $l$, the LHS of (22) is less or equal to the RHS of this equation. Since we also obtained that for large enough $l$, the LHS of (22) is larger than the RHS of this equation, the existence of a solution to (22) follows from continuity considerations.

## A.3 Proof of Inequality (30)

We prove here the inequality (30) (see p. 12) which implies that the "four-level quantization" lower bound on the parity-check density (see p. 10) is tighter than what can be interpreted as the "two levels quantization" bound in [13, Theorem 2.1]. Based on (2), we get

$$w = \Pr\{\text{LLR}(Y) < 0 \mid X = 0\} + \frac{1}{2}\Pr\{\text{LLR}(Y) = 0 \mid X = 0\}$$

so from (6), $w = p_2 + p_3$. By invoking Jensen's inequality, we get

$$\frac{(p_1 - p_2)^2}{p_1 + p_2} + \frac{(p_0 - p_3)^2}{p_0 + p_3}$$
$$= (p_1 + p_2)\left(\frac{p_1 - p_2}{p_1 + p_2}\right)^2 + (p_0 + p_3)\left(\frac{p_0 - p_3}{p_0 + p_3}\right)^2$$
$$\geq \left[(p_1 + p_2)\left(\frac{p_1 - p_2}{p_1 + p_2}\right) + (p_0 + p_3)\left(\frac{p_0 - p_3}{p_0 + p_3}\right)\right]^2$$
$$= (p_0 + p_1 - p_2 - p_3)^2$$
$$= (1 - 2p_2 - 2p_3)^2$$
$$= (1 - 2w)^2.$$

An equality is achieved if and only if $\frac{p_1 - p_2}{p_1 + p_2} = \frac{p_0 - p_3}{p_0 + p_3}$. From (A.7)–(A.10), we get

$$\frac{p_1 - p_2}{p_1 + p_2} = \frac{\int_{0^+}^{l} a(u)(1 - e^{-u})\, du}{\int_{0^+}^{l} a(u)(1 + e^{-u})\, du} \leq \frac{1 - e^{-l}}{1 + e^{-l}}$$

and

$$\frac{p_0 - p_3}{p_0 + p_3} = \frac{\int_{l}^{\infty} a(u)(1 - e^{-u})\, du}{\int_{l}^{\infty} a(u)(1 + e^{-u})\, du} \geq \frac{1 - e^{-l}}{1 + e^{-l}}.$$

The two fractions $\frac{p_1 - p_2}{p_1 + p_2}$ and $\frac{p_0 - p_3}{p_0 + p_3}$ cannot be equal unless the LLR is either equal to $l$ or $-l$. This makes the four-level quantization of the LLR identical to the two-level quantization used for the derivation of the original bound in [1, Theorem 2]. Equality can be also achieved if $p_1 + p_2 = 0$ or $p_0 + p_3 = 0$ which converts the channel model in Fig. 1 (see p. 6) to a BSC.

# Appendix B

## B.1 Proof of Lemma 3.3

Lemma 3.3 (see p. 16) is proved here by mathematical induction on the value of $k$.
$k = 1$: let $\tilde{s}$ denote the value of $s$ for which $k_s = 1$. In this case, we simply need to find the probability that the scalars $\Theta$ and $X$ differ. From (35)

$$\Pr\left\{\Theta = X \mid \boldsymbol{\Phi} = (a_1^{(\tilde{s})}, a_2^{(\tilde{s})}, \ldots, a_{d-1}^{(\tilde{s})})\right\}$$
$$= \frac{p_{\tilde{s}}}{p_{\tilde{s}} + p_{2^d-1-\tilde{s}}}$$
$$= \frac{1}{2}\left[1 + \left(1 - \frac{2p_{2^d-1-\tilde{s}}}{p_{\tilde{s}} + p_{2^d-1-\tilde{s}}}\right)\right]$$
$$= \frac{1}{2}\left[1 + \prod_{s=0}^{2^{d-1}-1}\left(1 - \frac{2p_{2^d-1-s}}{p_s + p_{2^d-1-s}}\right)^{k_s}\right].$$

Let us assume that for every $k < k'$ the claim holds, and prove it for $k = k'$. Let $\tilde{s}$ denote the value of $s$ for which $\boldsymbol{\Phi}_{i_1} = (a_1^{(\tilde{s})}, a_2^{(\tilde{s})}, \ldots, a_{d-1}^{(\tilde{s})})$. The probability that the components of the two random vectors $(\Theta_{i_1}, \Theta_{i_2}, \ldots, \Theta_{i_{k'}})$ and $(X_{i_1}, X_{i_2}, \ldots, X_{i_{k'}})$ differ in an even number of indices is equal to

$$q_1(\tilde{s})\, q_2(k', \tilde{s}) + \bigl(1 - q_1(\tilde{s})\bigr)\bigl(1 - q_2(k', \tilde{s})\bigr)$$

where $q_1(\tilde{s})$ designates the probability that $\Theta_{i_1} = X_{i_1}$, and $q_2(k', \tilde{s})$ designates the probability that the components of the two random vectors $(\Theta_{i_2}, \ldots, \Theta_{i_{k'}})$ and $(X_{i_2}, \ldots, X_{i_{k'}})$ differ in an even number of indices. Based on the assumption, we get

$$q_1(\tilde{s}) = \frac{1}{2}\left[1 + \left(1 - \frac{2p_{2^d-1-\tilde{s}}}{p_{\tilde{s}} + p_{2^d-1-\tilde{s}}}\right)\right]$$

$$q_2(k', \tilde{s}) = \frac{1}{2}\left[1 + \left(1 - \frac{2p_{2^d-1-\tilde{s}}}{p_{\tilde{s}} + p_{2^d-1-\tilde{s}}}\right)^{k_{\tilde{s}}-1} \cdot \prod_{\substack{s=0 \\ s \neq \tilde{s}}}^{2^{d-1}-1}\left(1 - \frac{2p_{2^d-1-s}}{p_s + p_{2^d-1-s}}\right)^{k_s}\right]$$

so the probability that $(\Theta_{i_1}, \ldots, \Theta_{i_{k'}})$ and $(X_{i_1}, \ldots, X_{i_{k'}})$ differ in an even number of indices is

$$q_1(\tilde{s})\, q_2(k', \tilde{s}) + \bigl(1 - q_1(\tilde{s})\bigr)\bigl(1 - q_2(k', \tilde{s})\bigr) = \frac{1}{2}\left[1 + \prod_{s=0}^{2^{d-1}-1}\left(1 - \frac{2p_{2^d-1-s}}{p_s + p_{2^d-1-s}}\right)^{k_s}\right].$$

This completes the proof of Lemma 3.3.

## B.2 Proof of the Property Claimed in the Discussion on Proposition 3.2

Following the discussion on Proposition 3.2 (see p. 17), we prove the existence of sub-optimal $2^{d+1}$ quantization levels, determined by $\tilde{l}_1, \ldots, \tilde{l}_{2^d-1}$ and their symmetric values around zero, so that even with this sub-optimal $2^{d+1}$-level quantization, the bound in the RHS of (34) is already tighter than the one which follows from the optimal choice of $2^d$ quantization levels. From the RHS of (34), it suffices to show that for any integer $k \geq 2$

$$\sum_{\substack{k_0,\ldots,k_{2^d-1} \\ \sum_i k_i = k}} \left\{ \binom{k}{k_0,\ldots,k_{2^d-1}} \prod_{i=0}^{2^d-1} (\tilde{p}_i + \tilde{p}_{2^{d+1}-1-i})^{k_i} \right.$$

$$\left. \cdot h_2 \left( \frac{1}{2} \left[ 1 - \prod_{i=0}^{2^d-1} \left( 1 - \frac{2\tilde{p}_{2^{d+1}-1-i}}{\tilde{p}_i + \tilde{p}_{2^{d+1}-1-i}} \right)^{k_i} \right] \right) \right\}$$

$$\leq \sum_{\substack{k_0,\ldots,k_{2^{d-1}-1} \\ \sum_i k_i = k}} \left\{ \binom{k}{k_0,\ldots,k_{2^{d-1}-1}} \prod_{i=0}^{2^{d-1}-1} \left( p_i^{(d)} + p_{2^d-1-i}^{(d)} \right)^{k_i} \right.$$

$$\left. \cdot h_2 \left( \frac{1}{2} \left[ 1 - \prod_{i=0}^{2^{d-1}-1} \left( 1 - \frac{2 p_{2^d-1-i}^{(d)}}{p_i^{(d)} + p_{2^d-1-i}^{(d)}} \right)^{k_i} \right] \right) \right\} \tag{B.1}$$

where $\tilde{p}_0, \tilde{p}_1, \ldots, \tilde{p}_{2^{d+1}-1}$ denote the transition probabilities, as defined in (33), which are associated with the above sub-optimal $2^{d+1}$ quantization levels. On the other hand, $p_0^{(d)}, p_1^{(d)}, \ldots, p_{2^d-1}^{(d)}$ denote the transition probabilities in (33) which correspond to the optimal $2^d$ quantization levels.

To prove (B.1), we define sub-optimal quantization levels $\tilde{l}_1, \ldots, \tilde{l}_{2^d-1}$ in the following way: For $i = 1, 2, \ldots, 2^{d-1} - 1$, we define $\tilde{l}_{2i} \triangleq l_i^{(d)}$, where $l_1^{(d)}, \ldots, l_{2^{d-1}-1}^{(d)}$ (and their symmetric values around zero) are the optimal $2^d$ quantization levels. The other levels (i.e., $\tilde{l}_j$ where the index $j$ is odd) are chosen arbitrarily as long as

$$\infty \triangleq \tilde{l}_0 \geq \tilde{l}_1 \geq \ldots \geq \tilde{l}_{2^d-1} \geq 0.$$

In a similar way to (33), let us the denote by $\tilde{p}_0, \ldots, \tilde{p}_{2^{d+1}-1}$ the transition probabilities associated with $\tilde{l}_1, \tilde{l}_2, \ldots, \tilde{l}_{2^d-1}$. This yields

$$\sum_{\substack{k_0,\ldots,k_{2^d-1} \\ \sum_i k_i = k}} \binom{k}{k_0,\ldots,k_{2^d-1}} \prod_{i=0}^{2^d-1} (\tilde{p}_i + \tilde{p}_{2^{d+1}-1-i})^{k_i} h_2 \left( \frac{1}{2} \left[ 1 - \prod_{i=0}^{2^d-1} \left( 1 - \frac{2\tilde{p}_{2^{d+1}-1-i}}{\tilde{p}_i + \tilde{p}_{2^{d+1}-1-i}} \right)^{k_i} \right] \right)$$

$$= \sum_{\substack{k_0,\ldots,k_{2^d-1} \\ \sum_i k_i = k}} \left\{ \binom{k}{k_0 + k_1, \ldots, k_{2^d-2} + k_{2^d-1}} \prod_{j=0}^{2^{d-1}-1} \binom{k_{2j} + k_{2j+1}}{k_{2j}} \right.$$

$$\left. \prod_{i=0}^{2^d-1} (\tilde{p}_i + \tilde{p}_{2^{d+1}-1-i})^{k_i} h_2 \left( \frac{1}{2} \left[ 1 - \prod_{i=0}^{2^d-1} \left( 1 - \frac{2\tilde{p}_{2^{d+1}-1-i}}{\tilde{p}_i + \tilde{p}_{2^{d+1}-1-i}} \right)^{k_i} \right] \right) \right\}.$$

Let us denote

$$k'_i = k_{2i} + k_{2i+1}, \quad i = 0, 1, \ldots, 2^{d-1} - 1$$

then, the above sum transforms to

$$\sum_{\substack{k'_0,\ldots,k'_{2^{d-1}-1} \\ \sum_i k'_i = k}} \left\{ \binom{k}{k'_0,\ldots,k'_{2^{d-1}-1}} \right.$$

$$\sum_{\substack{k_0,k_1,\ldots,k_{2^d-1} \\ \forall j:\ k_{2j}+k_{2j+1}=k'_j}} \prod_{j=0}^{2^{d-1}-1} \binom{k'_j}{k_{2j}} \left(\tilde{p}_{2j} + \tilde{p}_{2^{d+1}-1-2j}\right)^{k_{2j}} \left(\tilde{p}_{2j+1} + \tilde{p}_{2^{d+1}-1-(2j+1)}\right)^{k_{2j+1}}$$

$$\left. h_2\left(\frac{1}{2}\left[1 - \prod_{j=0}^{2^{d-1}-1}\left(1 - \frac{2\tilde{p}_{2^{d+1}-1-2j}}{\tilde{p}_{2j} + \tilde{p}_{2^{d+1}-1-2j}}\right)^{k_{2j}}\left(1 - \frac{2\tilde{p}_{2^{d+1}-1-(2j+1)}}{\tilde{p}_{2j+1} + \tilde{p}_{2^{d+1}-1-(2j+1)}}\right)^{k_{2j+1}}\right]\right)\right\}$$

$$= \sum_{\substack{k'_0,\ldots,k'_{2^{d-1}-1} \\ \sum_i k'_i = k}} \left\{ \binom{k}{k'_0,\ldots,k'_{2^{d-1}-1}} \prod_{j=0}^{2^{d-1}-1}\left(p^{(d)}_j + p^{(d)}_{2^d-1-j}\right)^{k'_j} \right.$$

$$\sum_{\substack{k_0,k_1,\ldots,k_{2^d-1} \\ \forall j:\ k_{2j}+k_{2j+1}=k'_j}} \prod_{j=0}^{2^{d-1}-1} \binom{k'_j}{k_{2j}} \frac{(\tilde{p}_{2j}+\tilde{p}_{2^{d+1}-1-2j})^{k_{2j}} (\tilde{p}_{2j+1}+\tilde{p}_{2^{d+1}-1-(2j+1)})^{k_{2j+1}}}{\left(p^{(d)}_j + p^{(d)}_{2^d-1-j}\right)^{k'_j}}$$

$$\left. h_2\left(\frac{1}{2}\left[1 - \prod_{j=0}^{2^{d-1}-1}\left(1 - \frac{2\tilde{p}_{2^{d+1}-1-2j}}{\tilde{p}_{2j} + \tilde{p}_{2^{d+1}-1-2j}}\right)^{k_{2j}}\left(1 - \frac{2\tilde{p}_{2^{d+1}-1-(2j+1)}}{\tilde{p}_{2j+1} + \tilde{p}_{2^{d+1}-1-(2j+1)}}\right)^{k_{2j+1}}\right]\right)\right\}.$$

(B.2)

Due to the choice of $\tilde{l}_j$, we have that $p^{(d)}_j = \tilde{p}_{2j} + \tilde{p}_{2j+1}$ for any $j = 0, 1, \ldots, 2^d - 1$. Hence,

$$\sum_{\substack{k_0,k_1,\ldots,k_{2^d-1} \\ \forall j:\ k_{2j}+k_{2j+1}=k'_j}} \prod_{j=0}^{2^{d-1}-1} \binom{k'_j}{k_{2j}} \frac{(\tilde{p}_{2j}+\tilde{p}_{2^{d+1}-1-2j})^{k_{2j}} (\tilde{p}_{2j+1}+\tilde{p}_{2^{d+1}-1-(2j+1)})^{k_{2j+1}}}{\left(p^{(d)}_j + p^{(d)}_{2^d-1-j}\right)^{k'_j}} \quad \text{(B.3)}$$

$$\stackrel{(a)}{=} \prod_{j=0}^{2^{d-1}-1} \sum_{\substack{k_{2j},k_{2j+1} \\ k_{2j}+k_{2j+1}=k'_j}} \binom{k'_j}{k_{2j}} \frac{(\tilde{p}_{2j}+\tilde{p}_{2^{d+1}-1-2j})^{k_{2j}} (\tilde{p}_{2j+1}+\tilde{p}_{2^{d+1}-1-(2j+1)})^{k_{2j+1}}}{\left(p^{(d)}_j + p^{(d)}_{2^d-1-j}\right)^{k'_j}}$$

$$= \prod_{j=0}^{2^{d-1}-1} \frac{\left((\tilde{p}_{2j}+\tilde{p}_{2j+1}) + (\tilde{p}_{2^{d+1}-1-2j}+\tilde{p}_{2^{d+1}-1-(2j+1)})\right)^{k'_j}}{\left(p^{(d)}_j + p^{(d)}_{2^d-1-j}\right)^{k'_j}}$$

$$= \prod_{j=0}^{2^{d-1}-1} \frac{\left(p^{(d)}_j + p^{(d)}_{2^d-1-j}\right)^{k'_j}}{\left(p^{(d)}_j + p^{(d)}_{2^d-1-j}\right)^{k'_j}} = 1$$

where the factorization, in (a), of the sum over the global function (whose variables are $k_0, k_1, \ldots, k_{2^d-1}$) into a product of sums of local functions (with variables $k_{2j}, k_{2j+1}$) follows from the concept of factor graphs. Therefore, the expression inside the sum in (B.3) forms a probability distribution.

Using the concavity of $h_2(\cdot)$, we apply Jensen's inequality to (B.2) (which is equal to the RHS of (B.1)) and get

$$\sum_{\substack{k_0,\ldots,k_{2^d-1} \\ \sum_i k_i = k}} \binom{k}{k_0,\ldots,k_{2^d-1}} \prod_{i=0}^{2^d-1} (\tilde{p}_i + \tilde{p}_{2^{d+1}-1-i})^{k_i} h_2\left(\frac{1}{2}\left[1 - \prod_{i=0}^{2^d-1}\left(1 - \frac{2\tilde{p}_{2^{d+1}-1-i}}{\tilde{p}_i + \tilde{p}_{2^{d+1}-1-i}}\right)^{k_i}\right]\right)$$

$$\leq \sum_{\substack{k'_0,\ldots,k'_{2^{d-1}-1} \\ \sum_i k'_i = k}} \binom{k}{k'_0,\ldots,k'_{2^{d-1}-1}} \prod_{j=0}^{2^{d-1}-1} \left(p^{(d)}_j + p^{(d)}_{2^d-1-j}\right)^{k'_j}$$

$$h_2\left(\frac{1}{2}\left[1 - \prod_{j=0}^{2^{d-1}-1} \sum_{\substack{k_{2j},k_{2j+1} \\ k_{2j}+k_{2j+1}=k'_j}} \binom{k'_j}{k_{2j}} \frac{(\tilde{p}_{2j} + \tilde{p}_{2^{d+1}-1-2j})^{k_{2j}}(\tilde{p}_{2j+1} + \tilde{p}_{2^{d+1}-1-(2j+1)})^{k_{2j+1}}}{\left(p^{(d)}_j + p^{(d)}_{2^d-1-j}\right)^{k'_j}}\right.\right.$$

$$\left.\left.\left(\frac{\tilde{p}_{2j} - \tilde{p}_{2^{d+1}-1-2j}}{\tilde{p}_{2j} + \tilde{p}_{2^{d+1}-1-2j}}\right)^{k_{2j}}\left(\frac{\tilde{p}_{2j+1} - \tilde{p}_{2^{d+1}-1-(2j+1)}}{\tilde{p}_{2j+1} + \tilde{p}_{2^{d+1}-1-(2j+1)}}\right)^{k_{2j+1}}\right]\right)$$

$$= \sum_{\substack{k'_0,\ldots,k'_{2^{d-1}-1} \\ \sum_i k'_i = k}} \binom{k}{k'_0,\ldots,k'_{2^{d-1}-1}} \prod_{j=0}^{2^{d-1}-1} \left(p^{(d)}_j + p^{(d)}_{2^d-1-j}\right)^{k'_j}$$

$$h_2\left(\frac{1}{2}\left[1 - \prod_{j=0}^{2^{d-1}-1} \sum_{\substack{k_{2j},k_{2j+1} \\ k_{2j}+k_{2j+1}=k'_j}} \binom{k'_j}{k_{2j}} \frac{(\tilde{p}_{2j} - \tilde{p}_{2^{d+1}-1-2j})^{k_{2j}}(\tilde{p}_{2j+1} - \tilde{p}_{2^{d+1}-1-(2j+1)})^{k_{2j+1}}}{\left(p^{(d)}_j + p^{(d)}_{2^d-1-j}\right)^{k'_j}}\right]\right)$$

$$= \sum_{\substack{k'_0,\ldots,k'_{2^{d-1}-1} \\ \sum_i k'_i = k}} \binom{k}{k'_0,\ldots,k'_{2^{d-1}-1}} \prod_{j=0}^{2^{d-1}-1} \left(p^{(d)}_j + p^{(d)}_{2^d-1-j}\right)^{k'_j}$$

$$h_2\left(\frac{1}{2}\left[1 - \prod_{j=0}^{2^{d-1}-1} \frac{\left((\tilde{p}_{2j} + \tilde{p}_{2j+1}) - (\tilde{p}_{2^{d+1}-1-2j} + \tilde{p}_{2^{d+1}-1-(2j+1)})\right)^{k'_j}}{\left(p^{(d)}_j + p^{(d)}_{2^d-1-j}\right)^{k'_j}}\right]\right)$$

$$= \sum_{\substack{k'_0,\ldots,k'_{2^{d-1}-1} \\ \sum_i k'_i = k}} \binom{k}{k'_0,\ldots,k'_{2^{d-1}-1}} \prod_{j=0}^{2^{d-1}-1} \left(p^{(d)}_j + p^{(d)}_{2^d-1-j}\right)^{k'_j} h_2\left(\frac{1}{2}\left[1 - \prod_{j=0}^{2^{d-1}-1} \frac{\left(p^{(d)}_j - p^{(d)}_{2^d-1-j}\right)^{k'_j}}{\left(p^{(d)}_j + p^{(d)}_{2^d-1-j}\right)^{k'_j}}\right]\right)$$

which therefore proves the sufficient condition for monotonicity, as stated in (B.1).

## Appendix C

We provide in this appendix further mathematical details related to the proof of Proposition 4.1. We note that Appendix C.2 serves here as a preparatory step for the derivation in Appendix C.3.

## C.1 Proof of Lemma 4.1

In order to prove Lemma 4.1, we first observe that

$$\Pr(\widetilde{\Theta} = 1 \mid \Omega = \alpha)$$
$$= \Pr(\Theta = -1 \mid \Omega = \alpha)$$
$$= \frac{a(-\alpha)}{a(\alpha) + a(-\alpha)}$$
$$= \frac{e^{-\alpha}}{1 + e^{-\alpha}}.$$

The proof of the lemma continues by mathematical induction. For $k = 1$, (56) holds since

$$\frac{1}{2}\left[1 - \left(1 - \frac{2e^{-\alpha_1}}{1 + e^{-\alpha_1}}\right)\right] = \frac{e^{-\alpha_1}}{1 + e^{-\alpha_1}} = \Pr(\widetilde{\Theta}_{i_1} = 1 \mid \Omega_{i_1} = \alpha_1).$$

Let us assume that (56) holds for $k' < k$, and prove it also for $k$. By our assumption and since the channel is memoryless, we get

$$\Pr\big(S_j = 1 \mid (\Omega_{i_1}, \ldots, \Omega_{i_k}) = (\alpha_1, \ldots, \alpha_k)\big)$$
$$= \Pr(\widetilde{\Theta}_{i_1} = 1 \mid \Omega_{i_1} = \alpha_1) \Pr\big(\text{Even number of ones in } (\Omega_{i_2}, \ldots, \Omega_{i_k})\big)$$
$$+ \Pr(\widetilde{\Theta}_{i_1} = 0 \mid \Omega_{i_1} = \alpha_1) \Pr\big(\text{Odd number of ones in } (\Omega_{i_2}, \ldots, \Omega_{i_k})\big)$$
$$\stackrel{(a)}{=} \frac{1}{2}\left[1 - \left(1 - \frac{2e^{-\alpha_1}}{1 + e^{-\alpha_1}}\right)\right] \cdot \left(1 - \frac{1}{2}\left[1 - \prod_{m=2}^{k}\left(1 - \frac{2e^{-\alpha_m}}{1 + e^{-\alpha_m}}\right)\right]\right)$$
$$+ \frac{1}{2}\left[1 + \left(1 - \frac{2e^{-\alpha_1}}{1 + e^{-\alpha_1}}\right)\right] \cdot \frac{1}{2}\left[1 - \prod_{m=2}^{k}\left(1 - \frac{2e^{-\alpha_m}}{1 + e^{-\alpha_m}}\right)\right]$$
$$= \frac{1}{2}\left[1 - \prod_{i=1}^{k}\left(1 - \frac{2e^{-\alpha_i}}{1 + e^{-\alpha_i}}\right)\right]$$

where equality (a) follows from the assumption that equation (56) holds for $k - 1$, and the proof follows by mathematical induction.

## C.2 Power Series Expansion of the Binary Entropy Function

**Lemma C.1.**
$$h_2(x) = 1 - \frac{1}{2\ln 2} \sum_{p=1}^{\infty} \frac{(1 - 2x)^{2p}}{p(2p - 1)}, \quad 0 \leq x \leq 1. \tag{C.1}$$

*Proof.* We prove this by expanding the binary entropy function into a power series around $\frac{1}{2}$. The first order derivative is

$$h_2'(x) = \frac{\ln\left(\frac{1-x}{x}\right)}{\ln 2}$$

and the higher order derivatives get the form

$$h_2^{(n)}(x) = -\frac{(n-2)!}{\ln 2}\left(\frac{(-1)^n}{x^{n-1}} + \frac{1}{(1-x)^{n-1}}\right), \quad n = 2, 3, \ldots.$$

The derivatives of odd degree therefore vanish at $x = \frac{1}{2}$, and for an even value of $n \geq 2$

$$h_2^{(n)}\left(\frac{1}{2}\right) = -\frac{(n-2)!\, 2^n}{\ln 2}.$$

This yields the following power series expansion of $h_2(\cdot)$ around $x = \frac{1}{2}$:

$$\begin{aligned}
h_2(x) &= 1 - \sum_{n \geq 2 \text{ even}} \left\{ \frac{\frac{(n-2)!\, 2^n}{\ln 2}}{n!} \cdot \left(x - \frac{1}{2}\right)^n \right\} \\
&= 1 - \frac{1}{\ln 2} \sum_{n \geq 2 \text{ even}} \frac{(2x-1)^n}{n(n-1)} \\
&= 1 - \frac{1}{2 \ln 2} \sum_{p=1}^{\infty} \frac{(2x-1)^{2p}}{p(2p-1)}
\end{aligned}$$

and this power series converges for all $x \in [0, 1]$. $\square$

We note that since the power series in (C.1) has always non-negative coefficients, then its truncation always gives an upper bound on the binary entropy function, i.e.,

$$h_2(x) \leq 1 - \frac{1}{2 \ln 2} \sum_{p=1}^{m} \frac{(1-2x)^{2p}}{p(2p-1)} \qquad \forall\, x \in [0, 1],\ m \in \mathbb{N}. \tag{C.2}$$

The case where $m = 1$ gives the upper bound in Lemma 3.2 which is used in this paper for the derivation of the lower bounds on the parity-check density. The reason for not using a tighter version of the binary entropy function for this case was because otherwise we would get a polynomial equation for $a_R$ whose solution cannot be given necessarily in closed form. As shown in Fig. 5, the upper bound on the binary entropy function, $h_2(\cdot)$, over the whole interval $[0, 1]$ is improved considerably by taking even a moderate value for $m$ (e.g., $m = 10$ gives already a very tight upper bound on $h_2(\cdot)$ which deviates from the exact values only at a small neighborhood near the two endpoints of this interval).

## C.3  Calculation of the Multi-Dimensional Integral in (57)

Based on Lemma C.1 which provides a power series expansion of $h_2(\cdot)$ near $\frac{1}{2}$, we obtain

$$\begin{aligned}
&\int_0^\infty \int_0^\infty \cdots \int_0^\infty \prod_{m=1}^k f_\Omega(\alpha_m)\, h_2\left(\frac{1}{2}\left(1 - \prod_{m=1}^k \left(\frac{1-e^{-\alpha_m}}{1+e^{-\alpha_m}}\right)\right)\right) d\alpha_1 d\alpha_2 \ldots d\alpha_k \\
&= 1 - \frac{1}{2\ln 2} \sum_{p=1}^{\infty} \frac{1}{p(2p-1)} \int_0^\infty \int_0^\infty \cdots \int_0^\infty \prod_{m=1}^k f_\Omega(\alpha_m) \prod_{m=1}^k \left(\frac{1-e^{-\alpha_m}}{1+e^{-\alpha_m}}\right)^{2p} d\alpha_1 d\alpha_2 \ldots d\alpha_k \\
&= 1 - \frac{1}{2\ln 2} \sum_{p=1}^{\infty} \frac{1}{p(2p-1)} \int_0^\infty \int_0^\infty \cdots \int_0^\infty \prod_{m=1}^k \left(f_\Omega(\alpha_m)\, \tanh^{2p}\left(\frac{\alpha_m}{2}\right)\right) d\alpha_1 d\alpha_2 \ldots d\alpha_k \\
&= 1 - \frac{1}{2\ln 2} \sum_{p=1}^{\infty} \frac{1}{p(2p-1)} \left(\int_0^\infty f_\Omega(\alpha)\, \tanh^{2p}\left(\frac{\alpha}{2}\right) d\alpha\right)^k.
\end{aligned}$$

This transforms the original $k$-dimensional integral to an infinite sum of one-dimensional integrals. Since we are interested to obtain a tight upper bound on the $k$-dimensional integral above, and since all the terms of the last infinite series are positive, then any truncation of the last infinite series is an upper bound. Based on the discussion in Appendix C.2, we will compute the first 10 terms of this series which (based on the plot in Fig. 5) will give a very tight upper bound on the $k$-dimensional integral (for all $k$). Hence, for the calculation of the un-quantized bounds on the thresholds of LDPC ensembles, we rely in our computations on the following tight upper bound on the $k$-dimensional integral for any integer $k$:

$$\int_0^\infty \int_0^\infty \ldots \int_0^\infty \prod_{m=1}^k f_w(\alpha_m)\, h_2\left(\frac{1}{2}\left(1 - \prod_{m=1}^k \left(\frac{1 - e^{-\alpha_m}}{1 + e^{-\alpha_m}}\right)\right)\right) d\alpha_1 d\alpha_2 \ldots d\alpha_k$$
$$\leq 1 - \frac{1}{2\ln 2} \sum_{p=1}^{10} \left\{\frac{1}{p(2p-1)} \left(\int_0^\infty f_w(\alpha)\, \tanh^{2p}\left(\frac{\alpha}{2}\right) d\alpha\right)^k\right\}.$$

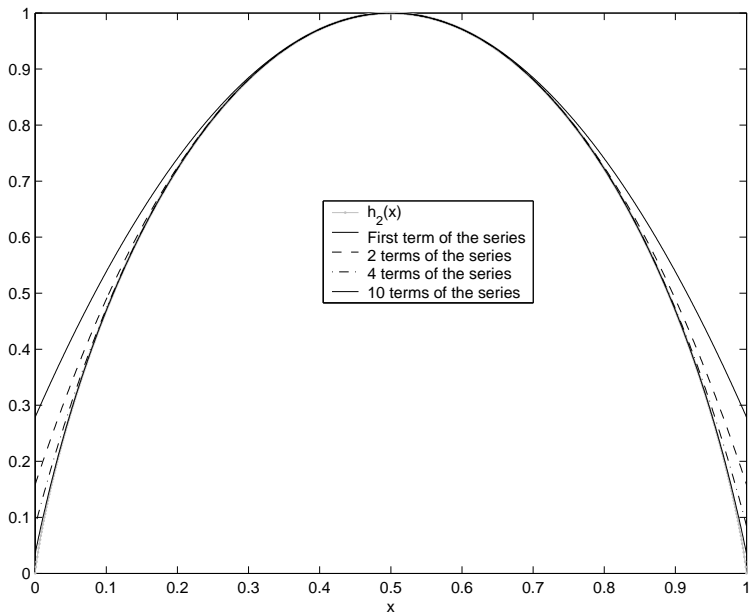

Figure 5: Plot of the binary entropy function to base 2 and some upper bounds which are obtained by truncating its power series around $x = \frac{1}{2}$.

This bound is clearly much tighter than the one we use in Section 4 for the derivation of a lower bound on the parity-check density, as for the derivation of the latter bound, we only take into account the first term (among the 10 positive terms) of the sum above.

# Appendix D

## D.1  Proof of Inequality (61)

Let $d \geq 2$ be an arbitrary integer and let $\infty \triangleq l_0 \geq l_1 \geq \ldots \geq l_{2^{d-1}-1} \geq 0$ (and their symmetric values around zero) be arbitrarily chosen quantization levels. To prove inequality (61), we start by applying the power series expansion of the binary entropy function (C.1) to the RHS of the

inequality and get

$$
\sum_k \left\{ d_k \sum_{\substack{k_0,\ldots,k_{2^{d-1}-1} \\ \sum_i k_i = k}} \binom{k}{k_0,\ldots,k_{2^{d-1}-1}} \cdot \prod_{i=0}^{2^{d-1}-1} (p_i + p_{2^d-1-i})^{k_i} h_2\left(\frac{1}{2}\left[1 - \prod_{i=0}^{2^{d-1}-1}\left(1 - \frac{2p_{2^d-1-i}}{p_i + p_{2^d-1-i}}\right)^{k_i}\right]\right) \right\}
$$

$$
= 1 - \frac{1}{2\ln(2)} \sum_{p=1}^{\infty} \left\{ \frac{1}{p(2p-1)} \sum_k \left\{ d_k \sum_{\substack{k_0,\ldots,k_{2^{d-1}-1} \\ \sum_i k_i = k}} \binom{k}{k_0,\ldots,k_{2^{d-1}-1}} \right.\right.
$$

$$
\left.\left. \cdot \prod_{i=0}^{2^{d-1}-1} (p_i + p_{2^d-1-i})^{k_i} \left(\frac{p_i - p_{2^d-1-i}}{p_i + p_{2^d-1-i}}\right)^{k_i \cdot 2p} \right\}\right\}.
$$

Therefore, a sufficient condition for (61) to hold, is that for any integers $k \geq 2$ and $p \geq 1$

$$
\left(\int_0^\infty a(l)(1+e^{-l})\tanh^{2p}\left(\frac{l}{2}\right) dl\right)^k
$$

$$
\geq \sum_{\substack{k_0,\ldots,k_{2^{d-1}-1} \\ \sum_i k_i = k}} \binom{k}{k_0,\ldots,k_{2^{d-1}-1}} \prod_{i=0}^{2^{d-1}-1} \left\{ (p_i + p_{2^d-1-i})^{k_i} \left(\frac{p_i - p_{2^d-1-i}}{p_i + p_{2^d-1-i}}\right)^{k_i \cdot 2p} \right\}. \quad \text{(D.1)}
$$

Let $f_\Omega(\cdot)$ be the *pdf* of the absolute value of the LLR, as defined in (51) (it is independent of the transmitted symbol $X$ because of the symmetry of the channel). Dividing the range of integration $[0,\infty]$ in the LHS of (D.1) into the $2^{d-1} - 1$ sub-intervals defined by the non-negative quantization levels $l_1, \ldots, l_{2^{d-1}-1}$ yields

$$
\left(\int_0^\infty a(l)(1+e^{-l})\tanh^{2p}\left(\frac{l}{2}\right) dl\right)^k
$$

$$
\overset{(a)}{=} \left(\int_0^\infty f_\Omega(l)\tanh^{2p}\left(\frac{l}{2}\right) dl\right)^k
$$

$$
= \left(\sum_{i=0}^{2^{d-1}-2} \underbrace{\int_{(l_{i+1})^+}^{l_i} f_\Omega(l)\tanh^{2p}\left(\frac{l}{2}\right) dl}_{\triangleq \varphi_i^{(p)}} + \underbrace{\int_0^{l_{2^{d-1}-1}} f_\Omega(l)\tanh^{2p}\left(\frac{l}{2}\right) dl}_{\triangleq \varphi_{2^{d-1}-1}^{(p)}} \right)^k
$$

$$
\triangleq \left(\sum_{i=0}^{2^{d-1}-1} \varphi_i^{(p)}\right)^k
$$

$$
= \sum_{\substack{k_0,\ldots,k_{2^{d-1}-1} \\ \sum_i k_i = k}} \binom{k}{k_0,\ldots,k_{2^{d-1}-1}} \prod_{i=0}^{2^{d-1}-1} \left(\varphi_i^{(p)}\right)^{k_i}
$$

where $(a)$ holds since $\tanh(0) = 0$. The above chain of equalities implies that in order to prove (D.1), it suffices to show that for any integer $p \geq 1$

$$
\varphi_i^{(p)} \geq (p_i + p_{2^d-1-i})\left(\frac{p_i - p_{2^d-1-i}}{p_i + p_{2^d-1-i}}\right)^{2p} \quad \forall\, i = 0, 1, \cdots, 2^{d-1} - 1. \quad \text{(D.2)}
$$

From (33), we have that

$$\int_{(l_{i+1})^+}^{l_i} f_\Omega(l) dl = p_i + p_{2^d-1-i} \quad \forall i = 0, \ldots, 2^{d-1} - 2$$

and

$$\int_0^{l_{2^{d-1}-1}} f_\Omega(l) dl = p_{2^{d-1}-1} + p_{2^{d-1}}$$

which implies that for every $i = 1, \ldots, 2^{d-1} - 1$, the function $\frac{f_\Omega(\cdot)}{p_i + p_{2^d-1-i}}$ is a *pdf* over the respective interval. Therefore, for $i \in \{1, \ldots, 2^{d-1} - 1\}$, we have

$$\begin{aligned}
\varphi_i^{(p)} &= (p_i + p_{2^d-1-i}) \int_{(l_{i+1})^+}^{l_i} \frac{f_\Omega(l)}{p_i + p_{2^d-1-i}} \left(\frac{1 - e^{-l}}{1 + e^{-l}}\right)^{2p} dl \\
&\stackrel{(a)}{\geq} (p_i + p_{2^d-1-i}) \left(\int_{(l_{i+1})^+}^{l_i} \frac{f_\Omega(l)}{p_i + p_{2^d-1-i}} \frac{1 - e^{-l}}{1 + e^{-l}} dl\right)^{2p} \\
&\stackrel{(b)}{=} (p_i + p_{2^d-1-i}) \left(\int_{(l_{i+1})^+}^{l_i} \frac{a(l)(1 - e^{-l})}{p_i + p_{2^d-1-i}} dl\right)^{2p} \\
&\stackrel{(c)}{=} (p_i + p_{2^d-1-i}) \left(\int_{(l_{i+1})^+}^{l_i} \frac{a(l) - a(-l)}{p_i + p_{2^d-1-i}} dl\right)^{2p} \\
&\stackrel{(d)}{=} (p_i + p_{2^d-1-i}) \left(\frac{p_i - p_{2^d-1-i}}{p_i + p_{2^d-1-i}}\right)^{2p} \quad \text{(D.3)}
\end{aligned}$$

where $(a)$ follows from the convexity of the function $f(x) = x^{2p}$ and by applying Jensen's equality, $(b)$ follows from the definition of $f_\Omega(\cdot)$ in (51), $(c)$ follows from the symmetry property of the *pdf* $a(\cdot)$, and $(d)$ follows from (33). For $i = 0$, the proof follows along the same lines as (D.3), except that in (b) we also use the fact that $\tanh(0) = 0$, so the two integrals from $0^+$ to $l_{2^{d-1}-1}$ and from $0$ to $l_{2^{d-1}-1}$ get the same value. This concludes the proof of the sufficient condition (D.2) for the satisfiability of inequality (61).

## D.2 Proof of the Claim Regarding the Tightness of the Lower Bound on the Parity-Check Density in Theorem 4.1

We show here that the lower bound on the parity-check density in Theorem 4.1 is uniformly tighter than the one in [13, Theorem 2.1] (except for the BSC and BEC where they coincide). In order to show this, we first prove the following lemma:

**Lemma D.1.** For any MBIOS channel, the following inequality holds

$$A \geq (1 - 2w)^2$$

where $w$ and $A$ are introduced in (2) and (64), respectively.

*Proof.* From (2), (51) and (64)

$$\begin{aligned}
A &= \int_0^\infty a(l) \, \frac{(1-e^{-l})^2}{1+e^{-l}} \, dl \\
&= \int_{0-}^\infty a(l) \, (1+e^{-l}) \, \tanh^2\left(\frac{l}{2}\right) dl \\
&= \int_{0-}^\infty f_\Omega(l) \, \tanh^2\left(\frac{l}{2}\right) dl \\
&\geq \left(\int_{0-}^\infty f_\Omega(l) \, \tanh\left(\frac{l}{2}\right) dl\right)^2 \\
&= \left(\int_{0-}^\infty a(l) \, (1+e^{-l}) \cdot \left(\frac{1-e^{-l}}{1+e^{-l}}\right) dl\right)^2 \\
&= \left(\int_{0-}^\infty a(l) \, dl - \int_{0-}^\infty e^{-l} \, a(l) \, dl\right)^2 \\
&= \left(\int_{0-}^\infty a(l) \, dl - \int_{0-}^\infty a(-l) \, dl\right)^2 \\
&= \left(\int_{0-}^\infty [a(l) + a(-l)] \, dl - 2\int_{0-}^\infty a(-l) \, dl\right)^2 \\
&= \left(1 + \Pr(\text{LLR} = 0) - 2\int_{0-}^\infty a(-l) \, dl\right)^2 \\
&= \left(1 + \Pr(\text{LLR} = 0) - 2\left(w + \frac{1}{2}\Pr(\text{LLR} = 0)\right)\right)^2 \\
&= (1-2w)^2.
\end{aligned}$$

where the single inequality above follows from Jensen's inequality. $\square$

The proof of our claim now follows directly by replacing the supremum over $x \in (0, A]$, which appears in the RHS of (62), with the same expression where we substitute $x = (1-2w)^2$.

### D.3 Proof for the Claim in Remark 4.4

In order to prove the claim in Remark 4.4 (see p. 27), it is required to show that

$$\frac{1-C}{1 - \frac{1}{2\ln(2)} \sum_{p=1}^\infty \left\{ \frac{1}{p(2p-1)} \sum_k d_k \left(\int_0^\infty a(l)\,(1+e^{-l})\,\tanh^{2p}\left(\frac{l}{2}\right) dl\right)^k \right\}} \geq \frac{2w}{1 - \sum_k d_k \, (1-2w)^k} \tag{D.4}$$

where $w$ is introduced in (2). The reason for showing this in light of the claim in Remark 4.4 is that the RHS of the last inequality follows from considerations related to a BEC, essentially in the same way that the second term of the maximization in the RHS of (49) is derived. By showing this, we prove that the maximization of the two expressions in the LHS and RHS of (D.4) doesn't affect the bound in Corollary 4.1.

Following the steps which lead to (68), we get that for any integer $k \geq 2$

$$\frac{1}{2\ln(2)} \sum_{p=1}^{\infty} \frac{1}{p(2p-1)} \left( \int_0^{\infty} a(l)(1+e^{-l}) \tanh^{2p}\left(\frac{l}{2}\right) dl \right)^k \geq C^k.$$

Applying this to (D.4) and denoting $\Omega(x) \triangleq \sum_k d_k x^k$, we get that a sufficient condition for (D.4) to hold is
$$\frac{1-C}{1-\Omega(C)} \geq \frac{2w}{1-\Omega(1-2w)}. \tag{D.5}$$

From the erasure decomposition lemma, we get that an MBIOS channel is physically degraded as compared to a BEC with an erasure probability $p = 2w$. By the information processing inequality, it follows that $C \leq 1-2w$. Therefore, in order to prove (D.5), it is enough to show that the function

$$f(x) = \frac{1-x}{1-\Omega(x)}$$

is monotonically decreasing for $x \in (0,1)$. We prove this property by showing that the derivative of $f(\cdot)$ is non-positive for $x \in (0,1)$. As the denominator of the derivative is positive, we may equivalently show
$$\Omega'(x)(1-x) - (1-\Omega(x)) \leq 0.$$

Dividing both sides of the inequality by $(1-x) \in (0,1)$ and noting that $\Omega(1) = \sum_k d_k = 1$, we get that it is enough to show
$$\Omega'(x) - \frac{\Omega(1) - \Omega(x)}{1-x} \leq 0. \tag{D.6}$$

Since $\Omega(\cdot)$ is a polynomial and therefore analytic, by the mean-value theorem we get that for some $\tilde{x} \in (x, 1)$
$$\frac{\Omega(1) - \Omega(x)}{1-x} = \Omega'(\tilde{x}).$$

Since $\Omega'(x) = \sum_k k d_k x^k$ is monotonically increasing for $x \geq 0$, then (D.6) follows for all $x \in (0,1)$. This in turn proves (D.4).

### D.4  Proof of Eq. (75)

In order to prove (75), we first multiply the two sides of (73) by $R$, and denote $R = (1-\varepsilon)C$. This gives that the lower bound on the bit error probability in (73) is non-positive if and only if
$$(1-C)B - \varepsilon C(1-B) \leq 0. \tag{D.7}$$

Unless the channel is noiseless, we get

$$\begin{aligned}
B &= \frac{1}{2\ln(2)} \sum_{p=1}^{\infty} \left\{ \frac{1}{p(2p-1)} \sum_k d_k \left( \int_{0+}^{\infty} a(l)(1+e^{-l}) \tanh^{2p}\left(\frac{l}{2}\right) dl \right)^k \right\} \\
&< \frac{1}{2\ln(2)} \sum_{p=1}^{\infty} \left\{ \frac{1}{p(2p-1)} \sum_k d_k \left( \int_{0+}^{\infty} a(l)(1+e^{-l}) dl \right)^k \right\} \\
&= \frac{1}{2\ln(2)} \sum_{p=1}^{\infty} \left\{ \frac{1}{p(2p-1)} \sum_k d_k \left( \int_{\mathbb{R}-\{0\}} a(l) dl \right)^k \right\} \\
&= \frac{1}{2\ln(2)} \sum_{p=1}^{\infty} \left\{ \frac{1}{p(2p-1)} \sum_k d_k \left( 1 - \Pr(\text{LLR}=0) \right)^k \right\} \\
&\leq \frac{1}{2\ln(2)} \sum_{p=1}^{\infty} \frac{1}{p(2p-1)} = 1.
\end{aligned}$$

Since $B < 1$, the LHS of (D.7) is monotonically decreasing in $\varepsilon$. We therefore deduce that the inequality (D.7) holds for $\varepsilon \geq \varepsilon_0$, where $\varepsilon_0$ is the solution of

$$(1-C)B - \varepsilon_0 C(1-B) = 0.$$

It can be readily seen that the solution of the last equation is given by $\varepsilon_0$ defined in (75).

### D.5 Proof of Eq. (76)

We will show both that there exists a unique $\varepsilon_0$ that satisfies (76), and that the RHS of (74) is non-positive if and only if $\varepsilon \geq \varepsilon_0$ where $\varepsilon_0$ is that unique solution. As in Appendix D.4, we begin by multiplying the two sides of (74) by $R$ and denoting $R = (1-\varepsilon)C$. It follows that the bound in the RHS of (74) is trivial (non-positive) if and only if

$$-\varepsilon C + \frac{1-(1-\varepsilon)C}{2\ln(2)} \sum_{p=1}^{\infty} \left\{ \frac{1}{p(2p-1)} \left( \int_0^{\infty} a(l)(1+e^{-l}) \tanh^{2p}\left(\frac{l}{2}\right) dl \right)^{\frac{(2-(1-\varepsilon)C)t}{1-(1-\varepsilon)C}} \right\} \leq 0. \quad (D.8)$$

We now show that the LHS of the last inequality is monotonically decreasing in $\varepsilon$. Let us denote

$$f(\varepsilon) \triangleq -\varepsilon C + \frac{1-(1-\varepsilon)C}{2\ln(2)} \sum_{p=1}^{\infty} \left\{ \frac{1}{p(2p-1)} \left( \int_0^{\infty} a(l)(1+e^{-l}) \tanh^{2p}\left(\frac{l}{2}\right) dl \right)^{\frac{(2-(1-\varepsilon)C)t}{1-(1-\varepsilon)C}} \right\}$$

$$\alpha_p \triangleq \frac{1}{2\ln(2)\, p(2p-1)}$$

$$a_p \triangleq \int_0^{\infty} a(l)(1+e^{-l}) \tanh^{2p}\left(\frac{l}{2}\right) dl.$$

By Dividing the derivative of $f$ w.r.t. $\varepsilon$ by $C$, we get

$$\begin{aligned}
\frac{f'(\varepsilon)}{C} &= \frac{1}{C}\Bigg( -C + C \sum_{p=1}^{\infty} \alpha_p\, a_p^{\frac{(2-(1-\varepsilon)C)t}{1-(1-\varepsilon)C}} \\
&\qquad + (1-(1-\varepsilon)C) \sum_{p=1}^{\infty} \alpha_p\, a_p^{\frac{(2-(1-\varepsilon)C)t}{1-(1-\varepsilon)C}} \log(a_p) \left( -\frac{tC}{(1-(1-\varepsilon)C)^2} \right) \Bigg) \\
&= \sum_{p=1}^{\infty} \left\{ \alpha_p \left( 1 - \log\left( a_p^{\frac{t}{1-(1-\varepsilon)C}} \right) \right) a_p^{\frac{(2-(1-\varepsilon)C)t}{1-(1-\varepsilon)C}} \right\} - 1.
\end{aligned}$$

From the symmetry property of $a(\cdot)$ and since $\tanh(x) \leq 1$ then $a_p \leq 1$, and it follows that $a_p^{\frac{(2-(1-\varepsilon)C)t}{1-(1-\varepsilon)C}} \leq a_p^{\frac{t}{1-(1-\varepsilon)C}}$. Therefore, the previous equality yields

$$\frac{f'(\varepsilon)}{C} \leq \sum_{p=1}^{\infty} \left\{ \alpha_p \left(1 - \log\left(a_p^{\frac{t}{1-(1-\varepsilon)C}}\right)\right) a_p^{\frac{t}{1-(1-\varepsilon)C}} \right\} - 1.$$

For $p \in \mathbb{N}$, let us denote $a_p^{\frac{t}{1-(1-\varepsilon)C}} \triangleq 1 - \delta_p$ where $0 < \delta_p < 1$. Therefore, we get from the previous inequality

$$\begin{aligned}
\frac{f'(\varepsilon)}{C} &\leq \sum_{p=1}^{\infty} \left\{ \alpha_p \left(1 - \log(1 - \delta_p)\right)(1 - \delta_p) \right\} - 1 \\
&\leq \sum_{p=1}^{\infty} \alpha_p - 1 = 0
\end{aligned}$$

where the second inequality follows from the inequality $\ln(1-x) > -\frac{x}{1-x}$ which holds for $x \in (0,1)$. This concludes the proof of the monotonicity of the LHS of (D.8). Observing that

$$f(0) = (1 - C) \sum_{p=1}^{\infty} \alpha_p \, a_p^{\frac{(2-C)t}{1-C}} > 0$$

and

$$\begin{aligned}
f(1) &= -C + \sum_{p=1}^{\infty} \alpha_p \, a_p^{2t} \\
&\leq -C + \sum_{p=1}^{\infty} \alpha_p \, a_p \\
&= -C + C = 0
\end{aligned}$$

where the first inequality follows since $a_p \leq 1$ and since $t \geq 1$ ($t = 1$ if and only if the code is cycle-free, otherwise $t > 1$.) The second equality follows from the last three equalities leading to (68). From the continuity of the function $f(\cdot)$ w.r.t. $\varepsilon$, we conclude that the monotonicity property of $f$, as shown above, ensures a unique solution for (76). From (D.8), it also follows from the monotonicity and continuity properties of $f(\cdot)$ in terms of $\varepsilon \in (0,1)$ that the RHS of (74) is non-positive if and only if $\varepsilon \geq \varepsilon_0$ where $\varepsilon_0$ is the unique solution of (76).

### Acknowledgment

The second author wishes to acknowledge Rüdiger Urbanke for stimulating discussions during the preparation of the work in [13] which also motivated the research in this paper.